\documentclass[aps,prl,twocolumn,superscriptaddress]{revtex4-1}
\usepackage{graphicx}
\usepackage{mathtools}
\usepackage{verbatim}
\usepackage{amsmath}
\usepackage{caption}
\usepackage{subcaption}
\usepackage{appendix}

\begin{document}

\title{Broadband Axion Dark Matter Haloscopes via Electric Sensing}

\author{Ben T. McAllister}
\email[]{ben.mcallister@uwa.edu.au}
\affiliation{ARC Centre of Excellence For Engineered Quantum Systems, Department of Physics, School of Physics and Mathematics, University of Western Australia, 35 Stirling Highway, Crawley WA 6009, Australia.}
\author{Maxim Goryachev}
\affiliation{ARC Centre of Excellence For Engineered Quantum Systems, Department of Physics, School of Physics and Mathematics, University of Western Australia, 35 Stirling Highway, Crawley WA 6009, Australia.}
\author{Jeremy Bourhill}
\affiliation{ARC Centre of Excellence For Engineered Quantum Systems, Department of Physics, School of Physics and Mathematics, University of Western Australia, 35 Stirling Highway, Crawley WA 6009, Australia.}
\author{Eugene N. Ivanov}
\affiliation{ARC Centre of Excellence For Engineered Quantum Systems, Department of Physics, School of Physics and Mathematics, University of Western Australia, 35 Stirling Highway, Crawley WA 6009, Australia.}
\author{Michael E. Tobar}
\email[]{michael.tobar@uwa.edu.au}
\affiliation{ARC Centre of Excellence For Engineered Quantum Systems, Department of Physics, School of Physics and Mathematics, University of Western Australia, 35 Stirling Highway, Crawley WA 6009, Australia.}

\date{\today}

\begin{abstract}
The mass of axion dark matter is only weakly bounded by cosmological observations, necessitating a variety of detection techniques over several orders of magnitude of mass ranges. Axions haloscopes based on resonant cavities have become the current standard to search for dark matter axions. Such structures are inherently narrowband and for low masses the volume of the required cavity becomes prohibitively large. Broadband low-mass detectors have already been proposed using inductive magnetometer sensors and a gapped toroidal solenoid magnet. In this work we propose an alternative, which uses electric sensors in a conventional solenoidal magnet aligned in the laboratory z-axis, as implemented in standard haloscope experiments. In the presence of the DC magnetic field, the inverse Primakoff effect causes a time varying permanent electric vacuum polarization in the z-direction to oscillate at the axion Compton frequency, which induces an oscillating electromotive force. We propose non-resonant techniques to detect this oscillating elctromotive force by implementing a capacitive sensor or an electric dipole antenna coupled to a low noise amplifier. We present the first experimental results and discuss the foundations and potential of this proposal. Preliminary results constrain $g_{a\gamma\gamma} >\sim2.35\times10^{-12}$ $\text{GeV}^{-1}$ in the mass range of $2.08\times10^{-11}$ to $2.2\times10^{-11}$ eV, and demonstrate potential sensitivity to axion-like dark matter with masses in the range of $10^{-12}$ to $10^{-8}$ eV.
\end{abstract}


\maketitle

For decades numerous cosmological observations have suggested the presence of a large amount of excess matter in the universe of unknown composition~\cite{RubinGRC,Bullet}. The lack of direct observation suggests the matter is only very weakly interacting with standard model particles - it is known as ``dark" matter. Many types of new particles have been proposed to account for the dark matter, over a vast mass range (sub-eV to GeV). Consequently we need a large number of experiments at various mass scales. Recent cosmological evidence combined with the null results of many experiments~\cite{NatureLowMass1,NatureLowMass2,XENON1T} has seen a resurgence of precision low-mass experiments. This work focuses on low mass axions or axion like particles (ALPs), which are hypothetical neutral, spin zero bosons often proposed to solve the strong charge-parity problem in QCD\cite{PQ1977,Wilczek1978,wisps}. Axions and ALPs can be formulated as dark matter~\cite{Sikivie1983b}, and if this is true they should be abundant in the laboratory frame on earth, and thus detectable. The most often explored ALP to standard model coupling is via the inverse Primakoff effect. In this coupling an axion interacts with a photon (usually a virtual photon supplied by a DC magnetic field) and converts into a second real photon such that;
\begin{equation*}
\hbar\omega_a\approx m_ac^2 + \frac{1}{2}m_av_a^2,
\end{equation*}
where $m_a$ is the mass of the axion, $\omega_a$ is the frequency of the generated real photon, $\hbar$ is the reduced Planck's constant, $c$ is the speed of light, and $v_a$ is the velocity of the axion with respect to the laboratory frame, the distribution of axion velocities with respect to earth gives a ``line-width" or effective quality-factor for the axion signal of approximately $10^6$~\cite{Sikivie83hal,Sikivie1985}. 

The strength of this axion-photon interaction, and the mass of the axion are given by,
\begin{equation*}
	\text{g}_{a\gamma\gamma}=\frac{\text{g}_{\gamma}\alpha}{f_{a}\pi},~~~~~~	
	\text{m}_a=\frac{z^{1/2}}{1+z}~\frac{f_{\pi}m_{\pi}}{f_a}.
	\label{eq:basic}
\end{equation*}
Here $z$ is the ratio of up and down quark masses, $\frac{m_u}{m_d}~\approx$ 0.56, $f_\pi$ is the pion decay constant $\approx$ 93 MeV, $m_\pi$ is the neutral pion mass $\approx$ 135 MeV, $\text{g}_\gamma$ is an axion-model dependent parameter of order 1, and $\alpha$ is the fine structure constant~\cite{K79,Kim2010,DFS81,SVZ80,Dine1983}. 
Confounding experimental efforts to detect axions via this coupling is the fact that $f_a$, the Peccei-Quinn Symmetry Breaking Scale, is unknown, and hence both the mass and strength of axion-photon coupling are unknown. This means that the photon frequency and amplitude of any axion induced signals are unknown, although we do have some broad limits from cosmological observations and previous experiments~\cite{Sikivie1983,Preskill1983}. There have also been theoretical predictions for axions over some specific mass ranges, but still includes broad range of masses~\cite{SMASH,MGU}. 

Most experiments that exploit the Primakoff effect rely on a tunable resonant structure designed to detect photons generated by axion conversion \cite{Sikivie83hal,ADMXaxions2010,MADMAX,YaleAxion,Cultask,ORGAN,DielectricPaper,Sikivie2014a,ADMX2011,McAllisterFormFactor,Garcon,BudkerPRX,Kim2014,PhysRevD.94.111702,HOANG2017,McAllister2016fux,Jooyoo2014}. The specific design depends heavily on the axion mass range, but most dark matter axion detection experiments operate in the radio frequency, microwave and millimeter-wave regimes. These experiments are inherently narrowband, which is a limitation, as the axion mass and therefore the corresponding photon frequency is unknown. More recently ABRACADABRA was proposed, which is partly a broadband low-mass particle haloscope~\cite{ABRACADABRA} designed to detect the photons generated by low mass, pre-inflationary dark matter axions. This experiment uses a solenoid magnet of gapped toroidal geometry and via the inverse Primakoff effect produces an oscillating magnetic field at the Compton frequency in the laboratory z-direction outside the solenoid. The oscillating magnetic field is detected by an inductive magnetometer sensor coil coupled to a SQUID amplifier. In this work we propose new techniques that may be implemented in the same setup as a standard resonant haloscope with a conventional solenoidal DC magnetic field aligned in the z-direction. We propose non-resonant techniques to detect this electric signal by considering capacitive sensors or electric dipole wire antennas coupled to low noise current and voltage amplifiers, which we show allows broadband sensitivity in a similar mass range to ABRACADABRA. We name this experiment: {\bfseries{B}}roadband {\bfseries{E}}lectric {\bfseries{A}}xion {\bfseries{S}}ensing {\bfseries{T}}echnique, or BEAST. 

\begin{figure}[t]
\includegraphics[width=\columnwidth]{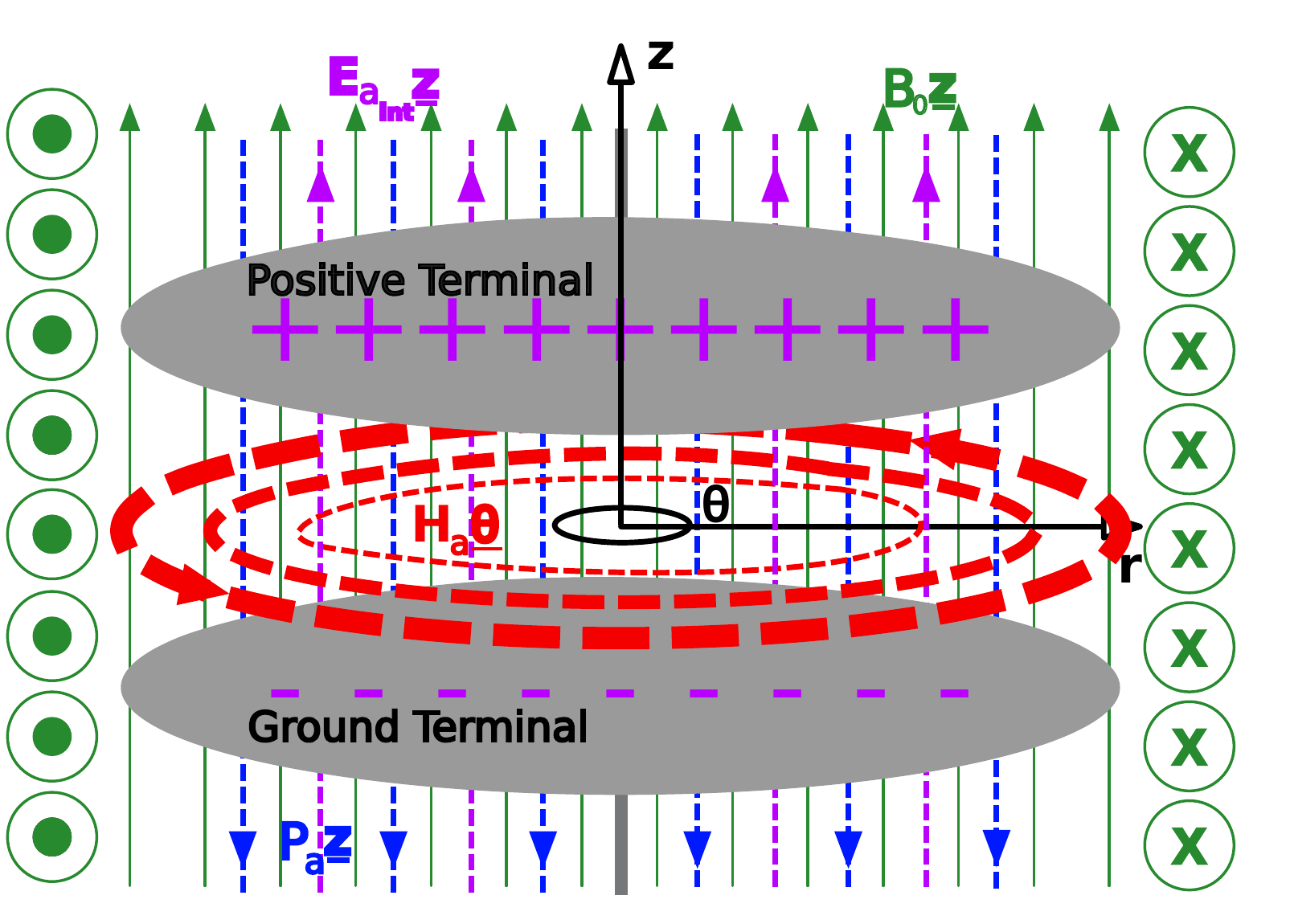}
\caption{A sketch of the proposed capacitor experiment, showing the the static magnetic field (green) and the axion-induced fields (blue and red), the induced EMF per unit length (purple) along with the alignment of the capacitor.}
\label{Diagram}
\end{figure}

Under a DC magnetic field, $\vec{B_0}$, the axion source term in the modified Maxwell's equations induced by the Primakoff effect can be represented as an effective current density of $\vec{J_a}= -g_{a\gamma\gamma}\sqrt{\frac{\epsilon_0}{\mu_0}}\vec{B_0}\frac{\partial{a}}{\partial{t}}$ (see~\cite{Jooyoo1990,HONG1991197,MTAxEd}). This current density is related to the effective axion charge density $\rho_a=g_{\alpha\gamma\gamma} c \vec{B_0}\cdot\nabla a,$ via a continuity equation $\vec{\nabla}\cdot \vec{J_a}=-\frac{\partial{\rho_a}}{\partial{t}}$.  Assuming there are no conductors involved, $\rho_a$ can be interpreted as an effective bound charge and $\vec{J_a}$ an effective polarization current associated with a permanent polarization, $\vec{P_a}= -\epsilon_0g_{a\gamma\gamma}a(c\vec{B_0})$  of the vacuum (as illustrated in Fig.\ref{Diagram}) \cite{MTAxEd}, consistent with Wilczek's original interpretation of the expectation value of this current from the modified Maxwell's equations \cite{Wilczek:1987aa}. This permanent polarization is actually a source term for an electromotive force per unit length given by $E_{a_{int}}=g_{a\gamma\gamma}a(cB_0)$\cite{MTAxEd}. For low-mass axions, this permanent polarization can either be detected directly with an electric sensor (as in this paper), or via the induced time-varying magnetic field using an inductive sensor, which is the basis of the ABRACADABRA experiment\cite{Sikivie2014a,ABRACADABRA}. It is apparent when designing an experiment sensitive to the induced electromotive force, the sensitivity is proportional to $a(t)$, while an experiment sensitive to the induced current density using a magnetic sensor is proportional to $\frac{\partial{a}}{\partial{t}}$.

In the case of a conductor such as a wire antenna, the electromotive force induces oscillating free charge, causing a free current induced by the inverse Primakoff effect, and one can likewise either detect the oscillating current through the wire or oscillating voltage across the antenna in the same way as the capacitance sensor (see~\cite{MTAxEd}). Both the wire and capacitor produce dipole fields, the main difference is the impedance supplied by the different electric sensors, for the capacitor it is reactive and for the wire antenna it is resistive.

First we considering a capacitor embedded in a solenoid such that the vector area of the plates are aligned with the applied DC $\vec{B}$-field in Fig.\ref{Diagram}. From the RF fields the expected voltage and current output from such a capacitor due to Primakoff axion conversion can be derived. The capacitance of a parallel plate capacitor with plate area $A$ and separation $d$ is given by $C=\frac{\epsilon_0\epsilon_r A}{d}$. Taking $a=a_0\cos(\omega_a t)$ and $a_0=\sqrt{\frac{2\rho_a}{c}}\frac{\hbar}{m_a}$~\cite{Daw:1998jm}, it can be shown that (see~\cite{MTAxEd}),
\begin{align}
\begin{split}
I_{a_{RMS}} &=  g_{a\gamma\gamma}A\sqrt{\frac{\epsilon_0}{\mu_0}}B_0\sqrt{\rho_ac^3}\\
V_{a_{RMS}} &=\frac{1}{\epsilon_r}g_{a\gamma\gamma} d\big(\frac{c}{\omega_a}\big)B_0 \sqrt{\rho_ac^3}.
\label{eq:RMS}
\end{split}
\end{align}
So, for typical parameters
\begin{align*}
I_{a_{RMS}} &= 6.43\times10^{-19}~Amperes \times \frac{g_{a\gamma\gamma}}{1.55\times10^{-18}~GeV^{-1}}\\
&\times\frac{A}{0.0079 ~m^2}\times\frac{B_0}{7~T}\times\sqrt{\frac{\rho_a}{0.45~\frac{GeV}{cm^3}}}\\
V_{a_{RMS}} &= 1.46\times10^{-13}~Volts \times\frac{g_{a\gamma\gamma}}{1.55\times10^{-18}~GeV^{-1}}\times\frac{1}{\epsilon_r}\\
&\times\frac{B_0}{7~T}\times\sqrt{\frac{\rho_a}{0.45~\frac{GeV}{cm^3}}}\times\frac{2\pi\times10^6~rad~s^{-1}}{\omega}\times\frac{d}{0.1~m}
\end{align*}
in Amperes and Volts rms respectively. Interestingly, neither the capacitor plate separation nor the permittivity between the plates has an impact on the expected current, however, the area of the capacitor plates is an important factor. Conversely, the voltage across the plates depends only on the permittivity and the plate separation, not the area. An optimal detector for such an experiment thus depends on the method of readout. If we wish to read current with, for example a SQUID, we would implement the largest diameter capacitors that would fit inside the magnet bore. If we were to read voltage directly with, for example, a high-impedance amplifier, we would design for the lowest permittivity and largest plate separation achievable. However, one must be mindful that the derivation assumes an ideal parallel plate structure, with no fringing or parasitic capacitance.

\begin{figure}[t!]
\includegraphics[width=1.0\columnwidth]{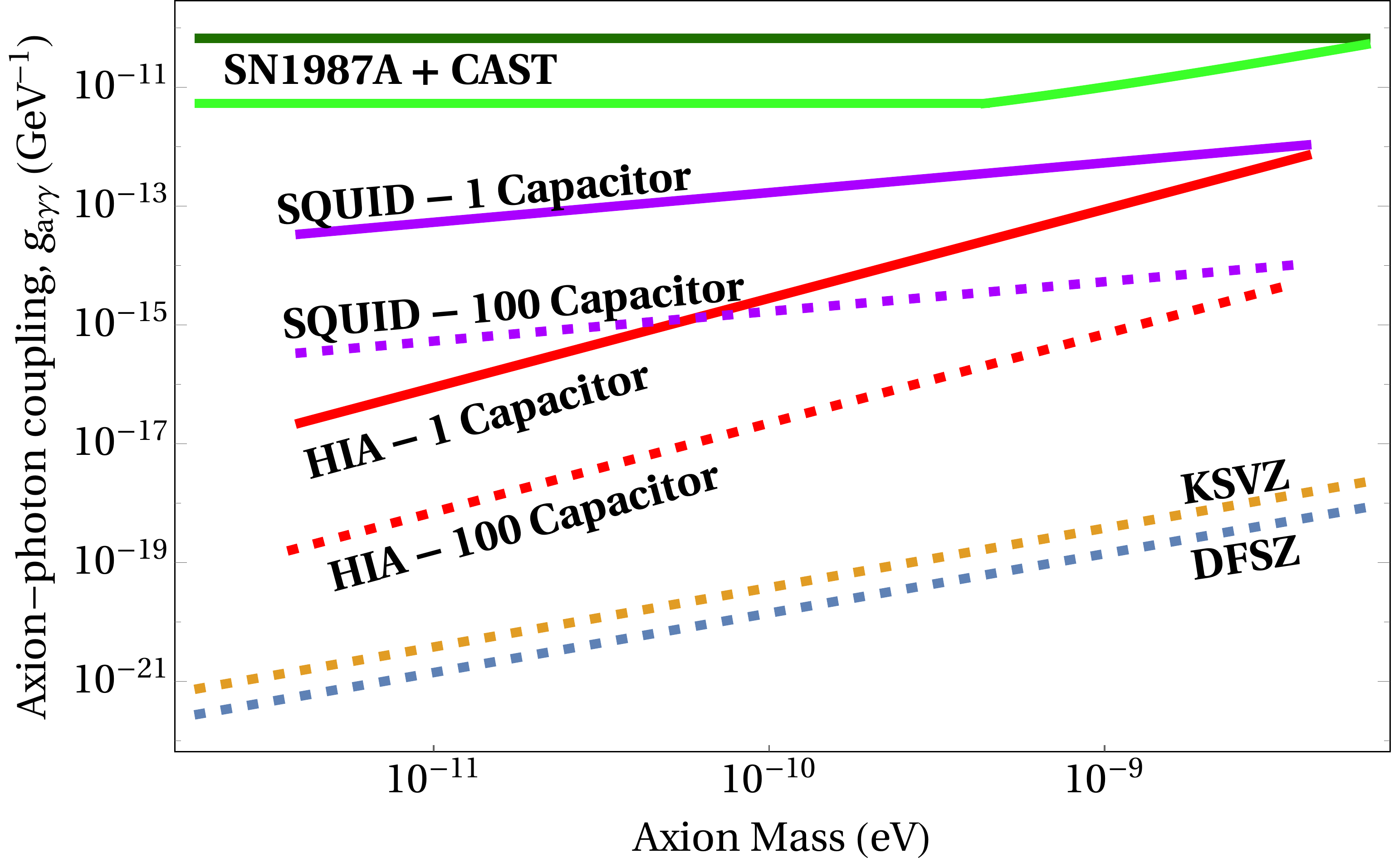}
\caption{Projected limits for the BEAST experiment, utilizing: a single capacitor (purple) and 100 capacitors (purple, dashed) coupled to a SQUID, and a single capacitor (red) and 100 capacitors (red, dashed) coupled to a high-impedance amplifier. Current best limits in the region from CAST (green) SN1987A (light green) are also plotted. Also shown are popular axion model bands, KSVZ (gold, dashed) and DFSZ (blue, dashed).}
\label{fig:projectedlimits}
\end{figure}

When compared with a typical resonant haloscope we lose the enhancement of the signal by the resonance quality-factor, Q. However, in principal it is possible to combine the signals from several capacitors in the same magnetic field to enhance sensitivity. Typical microwave Qs in axion haloscopes are of order $10^4-10^5$, and whilst it is unlikely that combining this many capacitors inside a single magnet bore would be readily achievable, it may be possible to mitigate the loss somewhat through combining capacitors. The optimal strategy for this capacitor combination depends on the method of readout, and various other limitations such as stray capacitance.

When implementing a voltage readout we note that the plate area is unimportant and may thus opt for a number of very small diameter capacitors with large plate separations. In such a case, to avoid issues associated with having very small plate areas with very large plate separations we may instead opt to combine many small capacitors (small plate area and small plate separation) in series to create a ``chain" of capacitors, with an effective total plate separation equal to the sum of the plate separations in the chain. This would maintain large effective plate separations, whilst simultaneously allowing for very low plate areas, and thus allowing for many such chains inside the same magnet. To understand how a scheme like this would work, cross capacitance between each element and stray capacitance due the experimental chamber and grounding would need to be modelled carefully. In contrast, implementing a current readout scheme, large plate areas are required with arbitrary plate separations. In such a scheme it would be optimal to create many small plate separation capacitors with large plate area and then combine the current outputs. Again placement within the magnet bore and the avoidance of stray capacitances or accidental electrical connection of neighbouring capacitors would need to be carefully considered. 

Another prospect for adding multiple electric sensors would be an array of wire dipole antennas aligned along the z-axis and to combine the outputs. Oscillating current in the wires would be driven by the Primakoff effect and these currents or resulting voltages could be combined in a similar way to the capacitor sensors discussed above. One benefit of this scheme is that the wire configuration should minimise the effect of stray capacitances. 

For the rest of this paper we focus on an axion haloscope sensitive to low-mass axions, where either a single or a number of capacitors are coupled to a low noise amplifier such as a SQUID or a high impedance amplifier inside a static magnetic field. This measurement is broadband and thus sensitive to axions over a wide range of mass values simultaneously. The bandwidth of the amplifier sets the limit on the detectable mass range. For example, in the case of a SQUID readout the location of the resonance which will be generated by the combination of the capacitor and SQUID input inductor (usually a few MHz)~\cite{SQUIDQuartz} will be the limiting factor. We now project the sensitivities for the experiments discussed above, with the two readout schemes presented and speculate on how to approach the axion model sensitivities.\\
\begin{figure}[t!]
\includegraphics[width=\columnwidth]{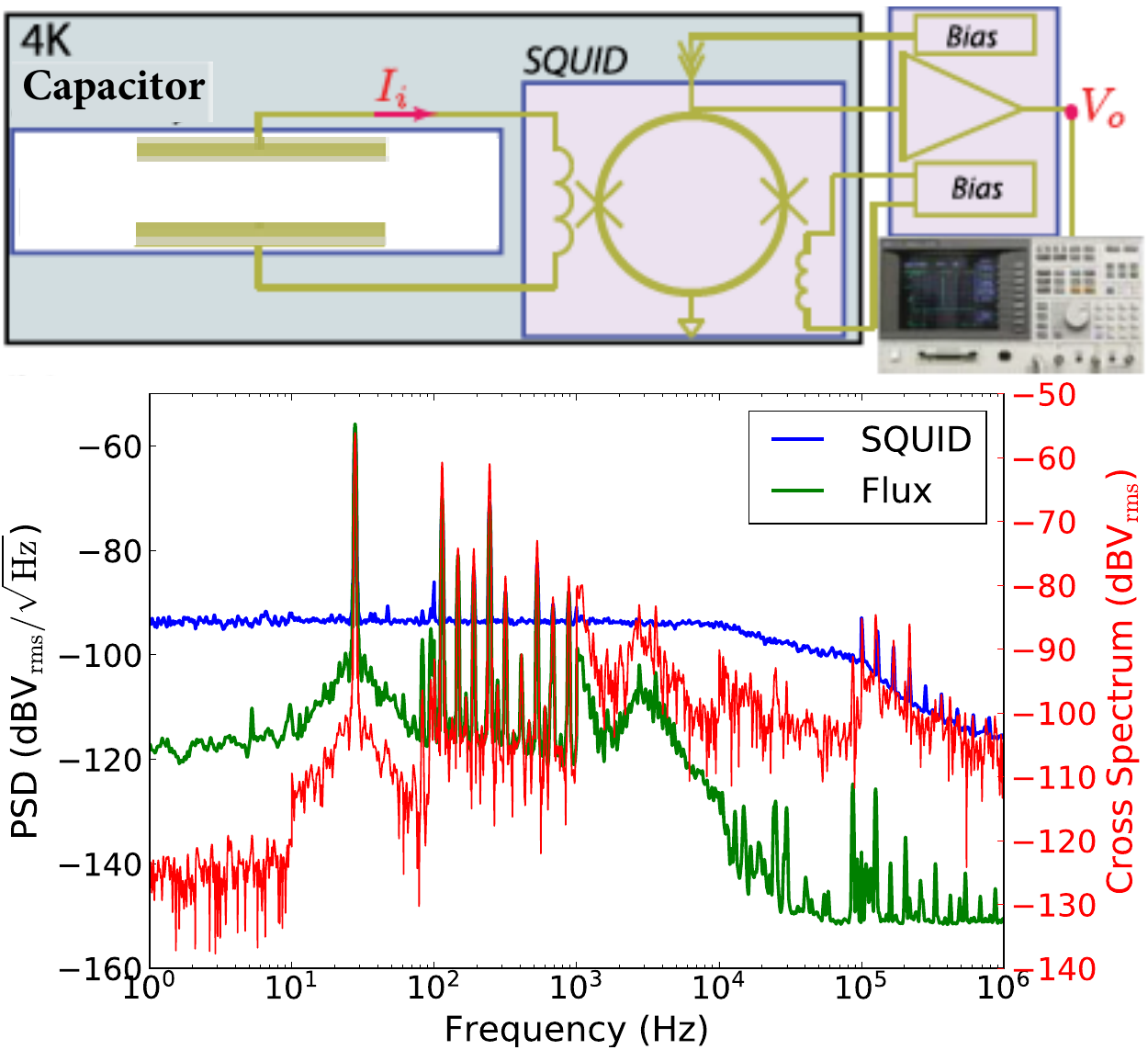}
\caption{Above: Schematic of the BEAST experiment with a capacitor coupled to a SQUID amplifier under a 7 Tesla DC field. Below: Voltage noise spectra at the output of the SQUID in the first run of BEAST. The blue (green) trace corresponds to the SQUID (flux) line, whilst the red trace corresponds to the cross-spectrum.}
\label{fig:widespectra}
\end{figure}
The peak of the spectral density of the RMS axion-induced current from the capacitor can be approximated by
\begin{equation}
\frac{I_{a_{RMS}}}{\sqrt{\pi\times\text{BW}}},
\label{eq:PSD}
\end{equation}
where BW is the bandwidth of the axion signal, roughly $10^{-6}$ times the central frequency. We can combine equations~\eqref{eq:RMS} and \eqref{eq:PSD} then compare with typical SQUID RMS current spectral density, $0.5~\frac{pA}{\sqrt{Hz}}$, to find the smallest possible $g_{a\gamma\gamma}$ detectable with such a setup as a function of axion mass. For general parameters this is given by,

\begin{equation*}
g_{a\gamma\gamma} = \frac{5.2\times 10^{-20}\sqrt{m}}{AB\sqrt{\rho}}.
\end{equation*}
With $m$ in eV, and everything else in SI units we obtain $g_{a\gamma\gamma}$ in $\text{GeV}^{-1}$. Fig.~\ref{fig:projectedlimits} shows projected limits for 10 cm diameter capacitors embedded in a 14 T magnetic field, coupled to SQUIDs, with arbitrary plate separation and an arbitrary material between the plates, assuming that the dark matter is comprised of axions with an energy density of 0.45 $\frac{\text{GeV}}{\text{cm}^3}$.

If we read out capacitor voltage with a high-impedance amplifier, we can follow a very similar process to estimate sensitivity. The effective total voltage noise referred to the input of high-impedance amplifiers depends on a number of factors, but for the purposes of this estimate we use a value of 100 $\frac{nV}{\sqrt{Hz}}$ (see appendix for calculations). Defining the peak spectral density of the RMS axion-induced voltage from the capacitor in the same way as for the current, we can arrive at the projected exclusion limits in much the same way. In this case we compare the peak of the axion-induced rms-voltage PSD with the above voltage noise. For general parameters we arrive at,
\begin{equation*}
g_{a\gamma\gamma} = \frac{4.4\times 10^{-10}m^{3/2}\epsilon_r}{Bd\sqrt{\rho}}.
\end{equation*}
With $m$ in eV, and everything else in SI units we obtain $g_{a\gamma\gamma}$ in $\text{GeV}^{-1}$. Fig.~\ref{fig:projectedlimits} also shows projected limits for capacitors (or chains of capacitors) with a net effective plate separation of 0.4 m, with vacuum between the plates, and assuming again that the dark matter is comprised of axions with an energy density of 0.45 $\frac{\text{GeV}}{\text{cm}^3}$.

We note that this detection method, due to its broadband nature, is preferable to traditional haloscopes in searching for transient enhancements in the flux of axions through the earth, such as those expected in the case of axion miniclusters~\cite{AxionMC}, or axion dark matter streaming~\cite{StreamingDM}. A traditional haloscope might miss such an event, due to being tuned away from the mass of the axion, whereas with this detector, if the axion mass falls within the bandwidth of the search we would be sensitive to these short-lived, but enormous boosts in sensitivity.\\
Our first BEAST experiment consists of a simple parallel plate capacitor coupled to a SQUID amplifier. In this experiment a $7.5\times 7$ cm rectangular capacitor was embedded in a 7 T field at 4 K for 8 days of observation time. The capacitor was coupled to a SQUID with a -3 dB bandwidth of 2.1 MHz and a transimpedance of 1.2 M$\Omega$. Fig.~\ref{fig:widespectra} shows spectra obtained up to 1 MHz with the associated spurious signals in this region. However, it is possible to discriminate against these spurious signals with the flux line, which is susceptible to spurious RF signals in the lab. These spectra do not have the requisite spectral resolution to resolve axion signals with an effective linewidth of $10^{-6}\times \omega_a$, but serve to demonstrate the expected output spectra of such an experiment and highlight the issue of spurious noise sources.

\begin{figure}[t!]
\includegraphics[width=0.5\textwidth]{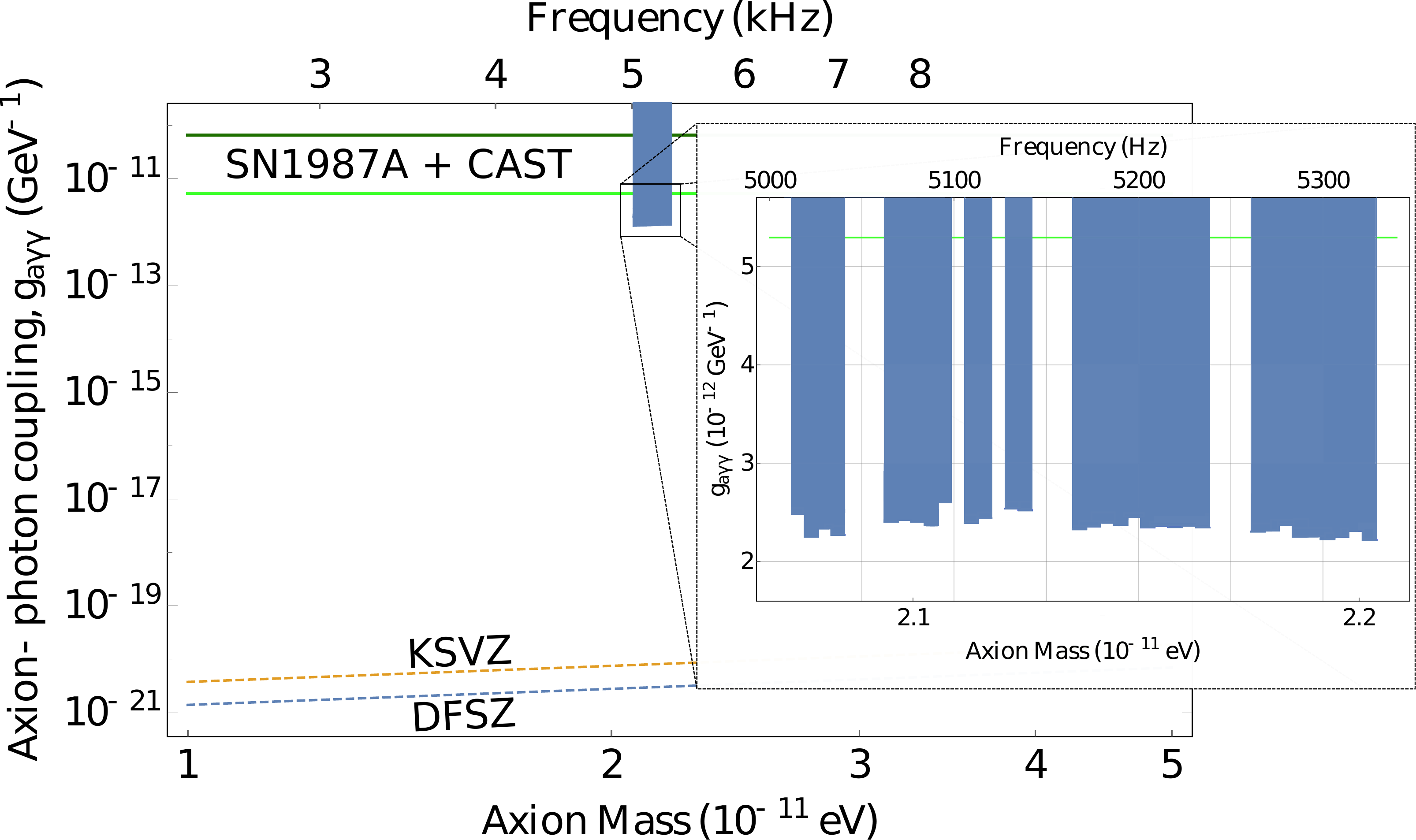}
\caption{Exclusion limits calculated from a single capacitor coupled to a SQUID. Previous best limits in the region from CAST (green) SN1987A (light green) are also plotted. Also shown are the axion model bands, KSVZ (gold, dashed) and DFSZ (blue, dashed). The inset shows the actual limit as a function of mass, including narrow regions where limits could not be placed due to large noise sources.}
\label{fig:limits}
\end{figure}

A higher-resolution search was conducted around 5 kHz, with the minimal spectral resolution of 4.5 mHz (increasing at higher frequencies), thus providing the requisite spectral resolution to detect ALP signals. All sharp peaks greater than $\sim$4.4 standard deviations from the mean originating from the SQUID were able to be excluded, due to a similar signal appearing in the flux line, as shown in Fig.~\ref{fig:widespectra}. Using this data, we may place the 95 \% confidence exclusion limits on axion-photon coupling shown in Fig.~\ref{fig:limits}. The average limit is $g_{a\gamma\gamma} > \sim2.35\times 10^{-12} GeV^{-1}$, with some variation as a function of mass. This experiment serves as a first result as well as a proof of concept, and can be readily extended to wider mass ranges. 

In conjunction with the planned ORGAN experiment~\cite{ORGAN}, we are developing the BEAST experiment for low-mass searches to run in parallel, utilizing the extra space in the ORGAN 14 T magnet bore. Technical limitations with availability of equipment, data acquisition and processing prevented a wider search from being feasible within the time scale of this first experiment, however in the future a dedicated system will be built and FPGA-based solutions to data acquisition issues will be implemented. Note, that if a direct voltage readout was employed via a high-impedance amplifier, we would not necessarily need to conduct the experiment cryogenically, as the noise floors of these devices are exceptionally low even at room temperature. Rare-earth magnets are capable of achieving magnetic fields on the order of a Tesla and, unlike superconducting solenoids, require no cryogenic environment to operate. Although these fields are considerably lower than those achievable with superconducting solenoids, an experiment could be conducted on a bench-top, without the need of a dedicated cryogenic cooler or magnet, and could operate continuously for very long times.

This work was funded by Australian Research Council grant No. DP160100253 and CE170100009, the Australian Government's Research Training Program, and the Bruce and Betty Green Foundation.

\appendix

\section{APPENDIX}

\subsection{Details of Input Voltage Noise of High-impedance amplifier}

In considering the effective voltage noise of the amplifier, referred to the input ($\delta u_{eff}$) we must consider contributions as a result of the input voltage noise ($\delta u_{V}$), the input current noise ($\delta u_{I}$), and the thermal input noise ($\delta u_{Th}$). This can be presented as,
\begin{equation*}
\delta u_{eff} = \sqrt{\delta u_{V}^2+\delta u_{I}^2+\delta u_{Th}^2}.
\end{equation*}
We will now discuss each of these quantities. $\delta u_{I}$ and $\delta u_{Th}$ must be found via,
\begin{align*}
\delta u_{I}=&\delta i_{amp}\times\left|Z\right|,\\
\delta u_{Th}=&\delta i_{Th}\times\left|Z\right|.
\end{align*}
Here $\delta i_{amp}$ is in the intrinsic amplifier current noise and $\delta i_{Th}$ is the thermally induced current noise given by
\begin{equation*}
\delta i_{Th} = \sqrt{\frac{k_B T_0}{R_{amp}}},
\end{equation*}
where $k_B$ is the Boltzmann constant, $T$ is the physical temperature, and $R_{amp}$ is the amplifier input resistance, and finally $Z$ is the complex impedance of the amplifier and capacitor system given by
\begin{equation*}
Z = \frac{R_{amp}}{1+i2\pi f C R_{amp}},
\end{equation*}
where $C$ takes into account the capacitance of the detector $C_{det}$ and the input capacitance of the amplifier $C_{amp}$ such that,
\begin{equation*}
C = \frac{C_{det}C_{amp}}{C_{det}+C_{amp}}.
\end{equation*}
The final quantity, $\delta u_{V}$ arises from the intrinsic amplifier input voltage noise $\delta u_{amp}$ according to
\begin{equation*}
\delta u_{V} = \delta u_{amp}\times\left|\frac{Z_{amp}}{Z_{amp}+Z_{det}}\right|.
\end{equation*}
Where $Z_{amp}$ and $Z_{det}$ are the amplifier and detector impedances given by
\begin{align*}
Z_{amp} =& \frac{R_{amp}}{1+i2\pi f C_{amp}R_{amp}},\\
Z_{det} =& \frac{1}{2\pi f C_{det}}.
\end{align*}
Taking the values for these equations from a suitable datasheet~\cite{HIA} and the detector parameters proposed in the main text, we arrive at a value for $\delta u_{eff}$ which is roughly constant as a function of frequency, at $\delta u_{eff} \approx 1\times10^{-7}\frac{V}{\sqrt{Hz}}$ from 1 kHz to 1 MHz. This will be heavily dependent on the specific parameters of the amplifier and detector, but for the purposes of these sensitivity estimates this is the value we employ.


\begin{thebibliography}{47}%
\makeatletter
\providecommand \@ifxundefined [1]{%
 \@ifx{#1\undefined}
}%
\providecommand \@ifnum [1]{%
 \ifnum #1\expandafter \@firstoftwo
 \else \expandafter \@secondoftwo
 \fi
}%
\providecommand \@ifx [1]{%
 \ifx #1\expandafter \@firstoftwo
 \else \expandafter \@secondoftwo
 \fi
}%
\providecommand \natexlab [1]{#1}%
\providecommand \enquote  [1]{``#1''}%
\providecommand \bibnamefont  [1]{#1}%
\providecommand \bibfnamefont [1]{#1}%
\providecommand \citenamefont [1]{#1}%
\providecommand \href@noop [0]{\@secondoftwo}%
\providecommand \href [0]{\begingroup \@sanitize@url \@href}%
\providecommand \@href[1]{\@@startlink{#1}\@@href}%
\providecommand \@@href[1]{\endgroup#1\@@endlink}%
\providecommand \@sanitize@url [0]{\catcode `\\12\catcode `\$12\catcode
  `\&12\catcode `\#12\catcode `\^12\catcode `\_12\catcode `\%12\relax}%
\providecommand \@@startlink[1]{}%
\providecommand \@@endlink[0]{}%
\providecommand \url  [0]{\begingroup\@sanitize@url \@url }%
\providecommand \@url [1]{\endgroup\@href {#1}{\urlprefix }}%
\providecommand \urlprefix  [0]{URL }%
\providecommand \Eprint [0]{\href }%
\providecommand \doibase [0]{http://dx.doi.org/}%
\providecommand \selectlanguage [0]{\@gobble}%
\providecommand \bibinfo  [0]{\@secondoftwo}%
\providecommand \bibfield  [0]{\@secondoftwo}%
\providecommand \translation [1]{[#1]}%
\providecommand \BibitemOpen [0]{}%
\providecommand \bibitemStop [0]{}%
\providecommand \bibitemNoStop [0]{.\EOS\space}%
\providecommand \EOS [0]{\spacefactor3000\relax}%
\providecommand \BibitemShut  [1]{\csname bibitem#1\endcsname}%
\let\auto@bib@innerbib\@empty
%</preamble>
\bibitem [{\citenamefont {{Rubin}}\ \emph {et~al.}(1980)\citenamefont
  {{Rubin}}, \citenamefont {{Ford}},\ and\ \citenamefont
  {{Thonnard}}}]{RubinGRC}%
  \BibitemOpen
  \bibfield  {author} {\bibinfo {author} {\bibfnamefont {V.~C.}\ \bibnamefont
  {{Rubin}}}, \bibinfo {author} {\bibfnamefont {W.~K.}\ \bibnamefont {{Ford}},
  \bibfnamefont {Jr.}}, \ and\ \bibinfo {author} {\bibfnamefont
  {N.}~\bibnamefont {{Thonnard}}},\ }\href {\doibase 10.1086/158003} {\bibfield
   {journal} {\bibinfo  {journal} {\apj}\ }\textbf {\bibinfo {volume} {238}},\
  \bibinfo {pages} {471} (\bibinfo {year} {1980})}\BibitemShut {NoStop}%
\bibitem [{\citenamefont {{Markevitch}}\ \emph {et~al.}(2004)\citenamefont
  {{Markevitch}}, \citenamefont {{Gonzalez}}, \citenamefont {{Clowe}},
  \citenamefont {{Vikhlinin}}, \citenamefont {{Forman}}, \citenamefont
  {{Jones}}, \citenamefont {{Murray}},\ and\ \citenamefont
  {{Tucker}}}]{Bullet}%
  \BibitemOpen
  \bibfield  {author} {\bibinfo {author} {\bibfnamefont {M.}~\bibnamefont
  {{Markevitch}}}, \bibinfo {author} {\bibfnamefont {A.~H.}\ \bibnamefont
  {{Gonzalez}}}, \bibinfo {author} {\bibfnamefont {D.}~\bibnamefont {{Clowe}}},
  \bibinfo {author} {\bibfnamefont {A.}~\bibnamefont {{Vikhlinin}}}, \bibinfo
  {author} {\bibfnamefont {W.}~\bibnamefont {{Forman}}}, \bibinfo {author}
  {\bibfnamefont {C.}~\bibnamefont {{Jones}}}, \bibinfo {author} {\bibfnamefont
  {S.}~\bibnamefont {{Murray}}}, \ and\ \bibinfo {author} {\bibfnamefont
  {W.}~\bibnamefont {{Tucker}}},\ }\href {\doibase 10.1086/383178} {\bibfield
  {journal} {\bibinfo  {journal} {\apj}\ }\textbf {\bibinfo {volume} {606}},\
  \bibinfo {pages} {819} (\bibinfo {year} {2004})},\ \Eprint
  {http://arxiv.org/abs/astro-ph/0309303} {astro-ph/0309303} \BibitemShut
  {NoStop}%
\bibitem [{\citenamefont {Barkana}(2018)}]{NatureLowMass1}%
  \BibitemOpen
  \bibfield  {author} {\bibinfo {author} {\bibfnamefont {R.}~\bibnamefont
  {Barkana}},\ }\bibfield  {booktitle} {\emph {\bibinfo {booktitle} {Nature}},\
  }\href@noop {} {\ \textbf {\bibinfo {volume} {555}},\ \bibinfo {pages} {71}
  (\bibinfo {year} {2018})}\BibitemShut {NoStop}%
\bibitem [{\citenamefont {Bowman}\ \emph {et~al.}(2018)\citenamefont {Bowman},
  \citenamefont {Rogers}, \citenamefont {Monsalve}, \citenamefont {Mozdzen},\
  and\ \citenamefont {Mahesh}}]{NatureLowMass2}%
  \BibitemOpen
  \bibfield  {author} {\bibinfo {author} {\bibfnamefont {J.~D.}\ \bibnamefont
  {Bowman}}, \bibinfo {author} {\bibfnamefont {A.~E.~E.}\ \bibnamefont
  {Rogers}}, \bibinfo {author} {\bibfnamefont {R.~A.}\ \bibnamefont
  {Monsalve}}, \bibinfo {author} {\bibfnamefont {T.~J.}\ \bibnamefont
  {Mozdzen}}, \ and\ \bibinfo {author} {\bibfnamefont {N.}~\bibnamefont
  {Mahesh}},\ }\bibfield  {booktitle} {\emph {\bibinfo {booktitle} {Nature}},\
  }\href@noop {} {\ \textbf {\bibinfo {volume} {555}},\ \bibinfo {pages} {67}
  (\bibinfo {year} {2018})}\BibitemShut {NoStop}%
\bibitem [{\citenamefont {Aprile}\ \emph {et~al.}(2017)\citenamefont {Aprile},
  \citenamefont {Aalbers}, \citenamefont {Agostini}, \citenamefont {Alfonsi},
  \citenamefont {Amaro}, \citenamefont {Anthony}, \citenamefont {Arneodo},
  \citenamefont {Barrow}, \citenamefont {Baudis}, \citenamefont {Bauermeister},
  \citenamefont {Benabderrahmane}, \citenamefont {Berger}, \citenamefont
  {Breur}, \citenamefont {Brown}, \citenamefont {Brown}, \citenamefont {Brown},
  \citenamefont {Bruenner}, \citenamefont {Bruno}, \citenamefont {Budnik},
  \citenamefont {B\"utikofer}, \citenamefont {Calv\'en}, \citenamefont
  {Cardoso}, \citenamefont {Cervantes}, \citenamefont {Cichon}, \citenamefont
  {Coderre}, \citenamefont {Colijn}, \citenamefont {Conrad}, \citenamefont
  {Cussonneau}, \citenamefont {Decowski}, \citenamefont {de~Perio},
  \citenamefont {Di~Gangi}, \citenamefont {Di~Giovanni}, \citenamefont
  {Diglio}, \citenamefont {Eurin}, \citenamefont {Fei}, \citenamefont
  {Ferella}, \citenamefont {Fieguth}, \citenamefont {Fulgione}, \citenamefont
  {Gallo~Rosso}, \citenamefont {Galloway}, \citenamefont {Gao}, \citenamefont
  {Garbini}, \citenamefont {Gardner}, \citenamefont {Geis}, \citenamefont
  {Goetzke}, \citenamefont {Grandi}, \citenamefont {Greene}, \citenamefont
  {Grignon}, \citenamefont {Hasterok}, \citenamefont {Hogenbirk}, \citenamefont
  {Howlett}, \citenamefont {Itay}, \citenamefont {Kaminsky}, \citenamefont
  {Kazama}, \citenamefont {Kessler}, \citenamefont {Kish}, \citenamefont
  {Landsman}, \citenamefont {Lang}, \citenamefont {Lellouch}, \citenamefont
  {Levinson}, \citenamefont {Lin}, \citenamefont {Lindemann}, \citenamefont
  {Lindner}, \citenamefont {Lombardi}, \citenamefont {Lopes}, \citenamefont
  {Manfredini}, \citenamefont {Maris}, \citenamefont {Marrod\'an~Undagoitia},
  \citenamefont {Masbou}, \citenamefont {Massoli}, \citenamefont {Masson},
  \citenamefont {Mayani}, \citenamefont {Messina}, \citenamefont {Micheneau},
  \citenamefont {Molinario}, \citenamefont {Mora}, \citenamefont {Murra},
  \citenamefont {Naganoma}, \citenamefont {Ni}, \citenamefont {Oberlack},
  \citenamefont {Pakarha}, \citenamefont {Pelssers}, \citenamefont {Persiani},
  \citenamefont {Piastra}, \citenamefont {Pienaar}, \citenamefont {Pizzella},
  \citenamefont {Piro}, \citenamefont {Plante}, \citenamefont {Priel},
  \citenamefont {Rauch}, \citenamefont {Reichard}, \citenamefont {Reuter},
  \citenamefont {Riedel}, \citenamefont {Rizzo}, \citenamefont {Rosendahl},
  \citenamefont {Rupp}, \citenamefont {Saldanha}, \citenamefont {dos Santos},
  \citenamefont {Sartorelli}, \citenamefont {Scheibelhut}, \citenamefont
  {Schindler}, \citenamefont {Schreiner}, \citenamefont {Schumann},
  \citenamefont {Scotto~Lavina}, \citenamefont {Selvi}, \citenamefont {Shagin},
  \citenamefont {Shockley}, \citenamefont {Silva}, \citenamefont {Simgen},
  \citenamefont {Sivers}, \citenamefont {Stein}, \citenamefont {Thapa},
  \citenamefont {Thers}, \citenamefont {Tiseni}, \citenamefont {Trinchero},
  \citenamefont {Tunnell}, \citenamefont {Vargas}, \citenamefont {Upole},
  \citenamefont {Wang}, \citenamefont {Wang}, \citenamefont {Wei},
  \citenamefont {Weinheimer}, \citenamefont {Wulf}, \citenamefont {Ye},
  \citenamefont {Zhang},\ and\ \citenamefont {Zhu}}]{XENON1T}%
  \BibitemOpen
  \bibfield  {author} {\bibinfo {author} {\bibfnamefont {E.}~\bibnamefont
  {Aprile}}, \bibinfo {author} {\bibfnamefont {J.}~\bibnamefont {Aalbers}},
  \bibinfo {author} {\bibfnamefont {F.}~\bibnamefont {Agostini}}, \bibinfo
  {author} {\bibfnamefont {M.}~\bibnamefont {Alfonsi}}, \bibinfo {author}
  {\bibfnamefont {F.~D.}\ \bibnamefont {Amaro}}, \bibinfo {author}
  {\bibfnamefont {M.}~\bibnamefont {Anthony}}, \bibinfo {author} {\bibfnamefont
  {F.}~\bibnamefont {Arneodo}}, \bibinfo {author} {\bibfnamefont
  {P.}~\bibnamefont {Barrow}}, \bibinfo {author} {\bibfnamefont
  {L.}~\bibnamefont {Baudis}}, \bibinfo {author} {\bibfnamefont
  {B.}~\bibnamefont {Bauermeister}}, \bibinfo {author} {\bibfnamefont {M.~L.}\
  \bibnamefont {Benabderrahmane}}, \bibinfo {author} {\bibfnamefont
  {T.}~\bibnamefont {Berger}}, \bibinfo {author} {\bibfnamefont {P.~A.}\
  \bibnamefont {Breur}}, \bibinfo {author} {\bibfnamefont {A.}~\bibnamefont
  {Brown}}, \bibinfo {author} {\bibfnamefont {A.}~\bibnamefont {Brown}},
  \bibinfo {author} {\bibfnamefont {E.}~\bibnamefont {Brown}}, \bibinfo
  {author} {\bibfnamefont {S.}~\bibnamefont {Bruenner}}, \bibinfo {author}
  {\bibfnamefont {G.}~\bibnamefont {Bruno}}, \bibinfo {author} {\bibfnamefont
  {R.}~\bibnamefont {Budnik}}, \bibinfo {author} {\bibfnamefont
  {L.}~\bibnamefont {B\"utikofer}}, \bibinfo {author} {\bibfnamefont
  {J.}~\bibnamefont {Calv\'en}}, \bibinfo {author} {\bibfnamefont {J.~M.~R.}\
  \bibnamefont {Cardoso}}, \bibinfo {author} {\bibfnamefont {M.}~\bibnamefont
  {Cervantes}}, \bibinfo {author} {\bibfnamefont {D.}~\bibnamefont {Cichon}},
  \bibinfo {author} {\bibfnamefont {D.}~\bibnamefont {Coderre}}, \bibinfo
  {author} {\bibfnamefont {A.~P.}\ \bibnamefont {Colijn}}, \bibinfo {author}
  {\bibfnamefont {J.}~\bibnamefont {Conrad}}, \bibinfo {author} {\bibfnamefont
  {J.~P.}\ \bibnamefont {Cussonneau}}, \bibinfo {author} {\bibfnamefont
  {M.~P.}\ \bibnamefont {Decowski}}, \bibinfo {author} {\bibfnamefont
  {P.}~\bibnamefont {de~Perio}}, \bibinfo {author} {\bibfnamefont
  {P.}~\bibnamefont {Di~Gangi}}, \bibinfo {author} {\bibfnamefont
  {A.}~\bibnamefont {Di~Giovanni}}, \bibinfo {author} {\bibfnamefont
  {S.}~\bibnamefont {Diglio}}, \bibinfo {author} {\bibfnamefont
  {G.}~\bibnamefont {Eurin}}, \bibinfo {author} {\bibfnamefont
  {J.}~\bibnamefont {Fei}}, \bibinfo {author} {\bibfnamefont {A.~D.}\
  \bibnamefont {Ferella}}, \bibinfo {author} {\bibfnamefont {A.}~\bibnamefont
  {Fieguth}}, \bibinfo {author} {\bibfnamefont {W.}~\bibnamefont {Fulgione}},
  \bibinfo {author} {\bibfnamefont {A.}~\bibnamefont {Gallo~Rosso}}, \bibinfo
  {author} {\bibfnamefont {M.}~\bibnamefont {Galloway}}, \bibinfo {author}
  {\bibfnamefont {F.}~\bibnamefont {Gao}}, \bibinfo {author} {\bibfnamefont
  {M.}~\bibnamefont {Garbini}}, \bibinfo {author} {\bibfnamefont
  {R.}~\bibnamefont {Gardner}}, \bibinfo {author} {\bibfnamefont
  {C.}~\bibnamefont {Geis}}, \bibinfo {author} {\bibfnamefont {L.~W.}\
  \bibnamefont {Goetzke}}, \bibinfo {author} {\bibfnamefont {L.}~\bibnamefont
  {Grandi}}, \bibinfo {author} {\bibfnamefont {Z.}~\bibnamefont {Greene}},
  \bibinfo {author} {\bibfnamefont {C.}~\bibnamefont {Grignon}}, \bibinfo
  {author} {\bibfnamefont {C.}~\bibnamefont {Hasterok}}, \bibinfo {author}
  {\bibfnamefont {E.}~\bibnamefont {Hogenbirk}}, \bibinfo {author}
  {\bibfnamefont {J.}~\bibnamefont {Howlett}}, \bibinfo {author} {\bibfnamefont
  {R.}~\bibnamefont {Itay}}, \bibinfo {author} {\bibfnamefont {B.}~\bibnamefont
  {Kaminsky}}, \bibinfo {author} {\bibfnamefont {S.}~\bibnamefont {Kazama}},
  \bibinfo {author} {\bibfnamefont {G.}~\bibnamefont {Kessler}}, \bibinfo
  {author} {\bibfnamefont {A.}~\bibnamefont {Kish}}, \bibinfo {author}
  {\bibfnamefont {H.}~\bibnamefont {Landsman}}, \bibinfo {author}
  {\bibfnamefont {R.~F.}\ \bibnamefont {Lang}}, \bibinfo {author}
  {\bibfnamefont {D.}~\bibnamefont {Lellouch}}, \bibinfo {author}
  {\bibfnamefont {L.}~\bibnamefont {Levinson}}, \bibinfo {author}
  {\bibfnamefont {Q.}~\bibnamefont {Lin}}, \bibinfo {author} {\bibfnamefont
  {S.}~\bibnamefont {Lindemann}}, \bibinfo {author} {\bibfnamefont
  {M.}~\bibnamefont {Lindner}}, \bibinfo {author} {\bibfnamefont
  {F.}~\bibnamefont {Lombardi}}, \bibinfo {author} {\bibfnamefont {J.~A.~M.}\
  \bibnamefont {Lopes}}, \bibinfo {author} {\bibfnamefont {A.}~\bibnamefont
  {Manfredini}}, \bibinfo {author} {\bibfnamefont {I.}~\bibnamefont {Maris}},
  \bibinfo {author} {\bibfnamefont {T.}~\bibnamefont {Marrod\'an~Undagoitia}},
  \bibinfo {author} {\bibfnamefont {J.}~\bibnamefont {Masbou}}, \bibinfo
  {author} {\bibfnamefont {F.~V.}\ \bibnamefont {Massoli}}, \bibinfo {author}
  {\bibfnamefont {D.}~\bibnamefont {Masson}}, \bibinfo {author} {\bibfnamefont
  {D.}~\bibnamefont {Mayani}}, \bibinfo {author} {\bibfnamefont
  {M.}~\bibnamefont {Messina}}, \bibinfo {author} {\bibfnamefont
  {K.}~\bibnamefont {Micheneau}}, \bibinfo {author} {\bibfnamefont
  {A.}~\bibnamefont {Molinario}}, \bibinfo {author} {\bibfnamefont
  {K.}~\bibnamefont {Mora}}, \bibinfo {author} {\bibfnamefont {M.}~\bibnamefont
  {Murra}}, \bibinfo {author} {\bibfnamefont {J.}~\bibnamefont {Naganoma}},
  \bibinfo {author} {\bibfnamefont {K.}~\bibnamefont {Ni}}, \bibinfo {author}
  {\bibfnamefont {U.}~\bibnamefont {Oberlack}}, \bibinfo {author}
  {\bibfnamefont {P.}~\bibnamefont {Pakarha}}, \bibinfo {author} {\bibfnamefont
  {B.}~\bibnamefont {Pelssers}}, \bibinfo {author} {\bibfnamefont
  {R.}~\bibnamefont {Persiani}}, \bibinfo {author} {\bibfnamefont
  {F.}~\bibnamefont {Piastra}}, \bibinfo {author} {\bibfnamefont
  {J.}~\bibnamefont {Pienaar}}, \bibinfo {author} {\bibfnamefont
  {V.}~\bibnamefont {Pizzella}}, \bibinfo {author} {\bibfnamefont {M.-C.}\
  \bibnamefont {Piro}}, \bibinfo {author} {\bibfnamefont {G.}~\bibnamefont
  {Plante}}, \bibinfo {author} {\bibfnamefont {N.}~\bibnamefont {Priel}},
  \bibinfo {author} {\bibfnamefont {L.}~\bibnamefont {Rauch}}, \bibinfo
  {author} {\bibfnamefont {S.}~\bibnamefont {Reichard}}, \bibinfo {author}
  {\bibfnamefont {C.}~\bibnamefont {Reuter}}, \bibinfo {author} {\bibfnamefont
  {B.}~\bibnamefont {Riedel}}, \bibinfo {author} {\bibfnamefont
  {A.}~\bibnamefont {Rizzo}}, \bibinfo {author} {\bibfnamefont
  {S.}~\bibnamefont {Rosendahl}}, \bibinfo {author} {\bibfnamefont
  {N.}~\bibnamefont {Rupp}}, \bibinfo {author} {\bibfnamefont {R.}~\bibnamefont
  {Saldanha}}, \bibinfo {author} {\bibfnamefont {J.~M.~F.}\ \bibnamefont {dos
  Santos}}, \bibinfo {author} {\bibfnamefont {G.}~\bibnamefont {Sartorelli}},
  \bibinfo {author} {\bibfnamefont {M.}~\bibnamefont {Scheibelhut}}, \bibinfo
  {author} {\bibfnamefont {S.}~\bibnamefont {Schindler}}, \bibinfo {author}
  {\bibfnamefont {J.}~\bibnamefont {Schreiner}}, \bibinfo {author}
  {\bibfnamefont {M.}~\bibnamefont {Schumann}}, \bibinfo {author}
  {\bibfnamefont {L.}~\bibnamefont {Scotto~Lavina}}, \bibinfo {author}
  {\bibfnamefont {M.}~\bibnamefont {Selvi}}, \bibinfo {author} {\bibfnamefont
  {P.}~\bibnamefont {Shagin}}, \bibinfo {author} {\bibfnamefont
  {E.}~\bibnamefont {Shockley}}, \bibinfo {author} {\bibfnamefont
  {M.}~\bibnamefont {Silva}}, \bibinfo {author} {\bibfnamefont
  {H.}~\bibnamefont {Simgen}}, \bibinfo {author} {\bibfnamefont {M.~v.}\
  \bibnamefont {Sivers}}, \bibinfo {author} {\bibfnamefont {A.}~\bibnamefont
  {Stein}}, \bibinfo {author} {\bibfnamefont {S.}~\bibnamefont {Thapa}},
  \bibinfo {author} {\bibfnamefont {D.}~\bibnamefont {Thers}}, \bibinfo
  {author} {\bibfnamefont {A.}~\bibnamefont {Tiseni}}, \bibinfo {author}
  {\bibfnamefont {G.}~\bibnamefont {Trinchero}}, \bibinfo {author}
  {\bibfnamefont {C.}~\bibnamefont {Tunnell}}, \bibinfo {author} {\bibfnamefont
  {M.}~\bibnamefont {Vargas}}, \bibinfo {author} {\bibfnamefont
  {N.}~\bibnamefont {Upole}}, \bibinfo {author} {\bibfnamefont
  {H.}~\bibnamefont {Wang}}, \bibinfo {author} {\bibfnamefont {Z.}~\bibnamefont
  {Wang}}, \bibinfo {author} {\bibfnamefont {Y.}~\bibnamefont {Wei}}, \bibinfo
  {author} {\bibfnamefont {C.}~\bibnamefont {Weinheimer}}, \bibinfo {author}
  {\bibfnamefont {J.}~\bibnamefont {Wulf}}, \bibinfo {author} {\bibfnamefont
  {J.}~\bibnamefont {Ye}}, \bibinfo {author} {\bibfnamefont {Y.}~\bibnamefont
  {Zhang}}, \ and\ \bibinfo {author} {\bibfnamefont {T.}~\bibnamefont {Zhu}}
  (\bibinfo {collaboration} {XENON Collaboration}),\ }\href {\doibase
  10.1103/PhysRevLett.119.181301} {\bibfield  {journal} {\bibinfo  {journal}
  {Phys. Rev. Lett.}\ }\textbf {\bibinfo {volume} {119}},\ \bibinfo {pages}
  {181301} (\bibinfo {year} {2017})}\BibitemShut {NoStop}%
\bibitem [{\citenamefont {Peccei}\ and\ \citenamefont {Quinn}(1977)}]{PQ1977}%
  \BibitemOpen
  \bibfield  {author} {\bibinfo {author} {\bibfnamefont {R.~D.}\ \bibnamefont
  {Peccei}}\ and\ \bibinfo {author} {\bibfnamefont {H.~R.}\ \bibnamefont
  {Quinn}},\ }\href {\doibase 10.1103/PhysRevLett.38.1440} {\bibfield
  {journal} {\bibinfo  {journal} {Phys. Rev. Lett.}\ }\textbf {\bibinfo
  {volume} {38}},\ \bibinfo {pages} {1440} (\bibinfo {year}
  {1977})}\BibitemShut {NoStop}%
\bibitem [{\citenamefont {Wilczek}(1978)}]{Wilczek1978}%
  \BibitemOpen
  \bibfield  {author} {\bibinfo {author} {\bibfnamefont {F.}~\bibnamefont
  {Wilczek}},\ }\href {\doibase 10.1103/PhysRevLett.40.279} {\bibfield
  {journal} {\bibinfo  {journal} {Phys. Rev. Lett.}\ }\textbf {\bibinfo
  {volume} {40}},\ \bibinfo {pages} {279} (\bibinfo {year} {1978})}\BibitemShut
  {NoStop}%
\bibitem [{\citenamefont {Jaeckel}\ and\ \citenamefont
  {Ringwald}(2010)}]{wisps}%
  \BibitemOpen
  \bibfield  {author} {\bibinfo {author} {\bibfnamefont {J.}~\bibnamefont
  {Jaeckel}}\ and\ \bibinfo {author} {\bibfnamefont {A.}~\bibnamefont
  {Ringwald}},\ }\href {\doibase 10.1146/annurev.nucl.012809.104433} {\bibfield
   {journal} {\bibinfo  {journal} {Annual Review of Nuclear and Particle
  Science}\ }\textbf {\bibinfo {volume} {60}},\ \bibinfo {pages} {405}
  (\bibinfo {year} {2010})}\BibitemShut {NoStop}%
\bibitem [{\citenamefont {Ipser}\ and\ \citenamefont
  {Sikivie}(1983)}]{Sikivie1983b}%
  \BibitemOpen
  \bibfield  {author} {\bibinfo {author} {\bibfnamefont {J.}~\bibnamefont
  {Ipser}}\ and\ \bibinfo {author} {\bibfnamefont {P.}~\bibnamefont
  {Sikivie}},\ }\href {\doibase 10.1103/PhysRevLett.50.925} {\bibfield
  {journal} {\bibinfo  {journal} {Phys. Rev. Lett.}\ }\textbf {\bibinfo
  {volume} {50}},\ \bibinfo {pages} {925} (\bibinfo {year} {1983})}\BibitemShut
  {NoStop}%
\bibitem [{\citenamefont {Sikivie}(1983)}]{Sikivie83hal}%
  \BibitemOpen
  \bibfield  {author} {\bibinfo {author} {\bibfnamefont {P.}~\bibnamefont
  {Sikivie}},\ }\href {\doibase 10.1103/PhysRevLett.51.1415} {\bibfield
  {journal} {\bibinfo  {journal} {Phys. Rev. Lett.}\ }\textbf {\bibinfo
  {volume} {51}},\ \bibinfo {pages} {1415} (\bibinfo {year}
  {1983})}\BibitemShut {NoStop}%
\bibitem [{\citenamefont {Sikivie}(1985)}]{Sikivie1985}%
  \BibitemOpen
  \bibfield  {author} {\bibinfo {author} {\bibfnamefont {P.}~\bibnamefont
  {Sikivie}},\ }\href {\doibase 10.1103/PhysRevD.32.2988} {\bibfield  {journal}
  {\bibinfo  {journal} {Phys. Rev. D}\ }\textbf {\bibinfo {volume} {32}},\
  \bibinfo {pages} {2988} (\bibinfo {year} {1985})}\BibitemShut {NoStop}%
\bibitem [{\citenamefont {Kim}(1979)}]{K79}%
  \BibitemOpen
  \bibfield  {author} {\bibinfo {author} {\bibfnamefont {J.~E.}\ \bibnamefont
  {Kim}},\ }\href {\doibase 10.1103/PhysRevLett.43.103} {\bibfield  {journal}
  {\bibinfo  {journal} {Phys. Rev. Lett.}\ }\textbf {\bibinfo {volume} {43}},\
  \bibinfo {pages} {103} (\bibinfo {year} {1979})}\BibitemShut {NoStop}%
\bibitem [{\citenamefont {Kim}\ and\ \citenamefont {Carosi}(2010)}]{Kim2010}%
  \BibitemOpen
  \bibfield  {author} {\bibinfo {author} {\bibfnamefont {J.~E.}\ \bibnamefont
  {Kim}}\ and\ \bibinfo {author} {\bibfnamefont {G.}~\bibnamefont {Carosi}},\
  }\href {\doibase 10.1103/RevModPhys.82.557} {\bibfield  {journal} {\bibinfo
  {journal} {Rev. Mod. Phys.}\ }\textbf {\bibinfo {volume} {82}},\ \bibinfo
  {pages} {557} (\bibinfo {year} {2010})}\BibitemShut {NoStop}%
\bibitem [{\citenamefont {Dine}\ \emph {et~al.}(1981)\citenamefont {Dine},
  \citenamefont {Fischler},\ and\ \citenamefont {Srednicki}}]{DFS81}%
  \BibitemOpen
  \bibfield  {author} {\bibinfo {author} {\bibfnamefont {M.}~\bibnamefont
  {Dine}}, \bibinfo {author} {\bibfnamefont {W.}~\bibnamefont {Fischler}}, \
  and\ \bibinfo {author} {\bibfnamefont {M.}~\bibnamefont {Srednicki}},\ }\href
  {\doibase http://dx.doi.org/10.1016/0370-2693(81)90590-6} {\bibfield
  {journal} {\bibinfo  {journal} {Physics Letters B}\ }\textbf {\bibinfo
  {volume} {104}},\ \bibinfo {pages} {199 } (\bibinfo {year}
  {1981})}\BibitemShut {NoStop}%
\bibitem [{\citenamefont {Shifman}\ \emph {et~al.}(1980)\citenamefont
  {Shifman}, \citenamefont {Vainshtein},\ and\ \citenamefont
  {Zakharov}}]{SVZ80}%
  \BibitemOpen
  \bibfield  {author} {\bibinfo {author} {\bibfnamefont {M.}~\bibnamefont
  {Shifman}}, \bibinfo {author} {\bibfnamefont {A.}~\bibnamefont {Vainshtein}},
  \ and\ \bibinfo {author} {\bibfnamefont {V.}~\bibnamefont {Zakharov}},\
  }\href {\doibase http://dx.doi.org/10.1016/0550-3213(80)90209-6} {\bibfield
  {journal} {\bibinfo  {journal} {Nuclear Physics B}\ }\textbf {\bibinfo
  {volume} {166}},\ \bibinfo {pages} {493 } (\bibinfo {year}
  {1980})}\BibitemShut {NoStop}%
\bibitem [{\citenamefont {Dine}\ and\ \citenamefont
  {Fischler}(1983)}]{Dine1983}%
  \BibitemOpen
  \bibfield  {author} {\bibinfo {author} {\bibfnamefont {M.}~\bibnamefont
  {Dine}}\ and\ \bibinfo {author} {\bibfnamefont {W.}~\bibnamefont
  {Fischler}},\ }\href {\doibase
  http://dx.doi.org/10.1016/0370-2693(83)90639-1} {\bibfield  {journal}
  {\bibinfo  {journal} {Physics Letters B}\ }\textbf {\bibinfo {volume}
  {120}},\ \bibinfo {pages} {137 } (\bibinfo {year} {1983})}\BibitemShut
  {NoStop}%
\bibitem [{\citenamefont {Abbott}\ and\ \citenamefont
  {Sikivie}(1983)}]{Sikivie1983}%
  \BibitemOpen
  \bibfield  {author} {\bibinfo {author} {\bibfnamefont {L.}~\bibnamefont
  {Abbott}}\ and\ \bibinfo {author} {\bibfnamefont {P.}~\bibnamefont
  {Sikivie}},\ }\href {\doibase http://dx.doi.org/10.1016/0370-2693(83)90638-X}
  {\bibfield  {journal} {\bibinfo  {journal} {Physics Letters B}\ }\textbf
  {\bibinfo {volume} {120}},\ \bibinfo {pages} {133 } (\bibinfo {year}
  {1983})}\BibitemShut {NoStop}%
\bibitem [{\citenamefont {Preskill}\ \emph {et~al.}(1983)\citenamefont
  {Preskill}, \citenamefont {Wise},\ and\ \citenamefont
  {Wilczek}}]{Preskill1983}%
  \BibitemOpen
  \bibfield  {author} {\bibinfo {author} {\bibfnamefont {J.}~\bibnamefont
  {Preskill}}, \bibinfo {author} {\bibfnamefont {M.~B.}\ \bibnamefont {Wise}},
  \ and\ \bibinfo {author} {\bibfnamefont {F.}~\bibnamefont {Wilczek}},\ }\href
  {\doibase http://dx.doi.org/10.1016/0370-2693(83)90637-8} {\bibfield
  {journal} {\bibinfo  {journal} {Physics Letters B}\ }\textbf {\bibinfo
  {volume} {120}},\ \bibinfo {pages} {127 } (\bibinfo {year}
  {1983})}\BibitemShut {NoStop}%
\bibitem [{\citenamefont {Ballesteros}\ \emph {et~al.}(2017)\citenamefont
  {Ballesteros}, \citenamefont {Redondo}, \citenamefont {Ringwald},\ and\
  \citenamefont {Tamarit}}]{SMASH}%
  \BibitemOpen
  \bibfield  {author} {\bibinfo {author} {\bibfnamefont {G.}~\bibnamefont
  {Ballesteros}}, \bibinfo {author} {\bibfnamefont {J.}~\bibnamefont
  {Redondo}}, \bibinfo {author} {\bibfnamefont {A.}~\bibnamefont {Ringwald}}, \
  and\ \bibinfo {author} {\bibfnamefont {C.}~\bibnamefont {Tamarit}},\ }\href
  {\doibase 10.1103/PhysRevLett.118.071802} {\bibfield  {journal} {\bibinfo
  {journal} {Phys. Rev. Lett.}\ }\textbf {\bibinfo {volume} {118}},\ \bibinfo
  {pages} {071802} (\bibinfo {year} {2017})},\ \Eprint
  {http://arxiv.org/abs/1608.05414} {arXiv:1608.05414 [hep-ph]} \BibitemShut
  {NoStop}%
%%CITATION = ARXIV:1608.05414;%%
\bibitem [{\citenamefont {Luzio}\ \emph {et~al.}(2018)\citenamefont {Luzio},
  \citenamefont {Ringwald},\ and\ \citenamefont {Tamarit}}]{MGU}%
  \BibitemOpen
  \bibfield  {author} {\bibinfo {author} {\bibfnamefont {L.~D.}\ \bibnamefont
  {Luzio}}, \bibinfo {author} {\bibfnamefont {A.}~\bibnamefont {Ringwald}}, \
  and\ \bibinfo {author} {\bibfnamefont {C.}~\bibnamefont {Tamarit}},\
  }\href@noop {} {\bibfield  {journal} {\bibinfo  {journal} {arXiv:1807.09769v1
  [hep-ph]}\ } (\bibinfo {year} {2018})}\BibitemShut {NoStop}%
\bibitem [{\citenamefont {Asztalos}\ \emph {et~al.}(2010)\citenamefont
  {Asztalos}, \citenamefont {Carosi}, \citenamefont {Hagmann}, \citenamefont
  {Kinion}, \citenamefont {van Bibber}, \citenamefont {Hotz}, \citenamefont
  {Rosenberg}, \citenamefont {Rybka}, \citenamefont {Hoskins}, \citenamefont
  {Hwang}, \citenamefont {Sikivie}, \citenamefont {Tanner}, \citenamefont
  {Bradley},\ and\ \citenamefont {Clarke}}]{ADMXaxions2010}%
  \BibitemOpen
  \bibfield  {author} {\bibinfo {author} {\bibfnamefont {S.~J.}\ \bibnamefont
  {Asztalos}}, \bibinfo {author} {\bibfnamefont {G.}~\bibnamefont {Carosi}},
  \bibinfo {author} {\bibfnamefont {C.}~\bibnamefont {Hagmann}}, \bibinfo
  {author} {\bibfnamefont {D.}~\bibnamefont {Kinion}}, \bibinfo {author}
  {\bibfnamefont {K.}~\bibnamefont {van Bibber}}, \bibinfo {author}
  {\bibfnamefont {M.}~\bibnamefont {Hotz}}, \bibinfo {author} {\bibfnamefont
  {L.~J.}\ \bibnamefont {Rosenberg}}, \bibinfo {author} {\bibfnamefont
  {G.}~\bibnamefont {Rybka}}, \bibinfo {author} {\bibfnamefont
  {J.}~\bibnamefont {Hoskins}}, \bibinfo {author} {\bibfnamefont
  {J.}~\bibnamefont {Hwang}}, \bibinfo {author} {\bibfnamefont
  {P.}~\bibnamefont {Sikivie}}, \bibinfo {author} {\bibfnamefont {D.~B.}\
  \bibnamefont {Tanner}}, \bibinfo {author} {\bibfnamefont {R.}~\bibnamefont
  {Bradley}}, \ and\ \bibinfo {author} {\bibfnamefont {J.}~\bibnamefont
  {Clarke}},\ }\href {\doibase 10.1103/PhysRevLett.104.041301} {\bibfield
  {journal} {\bibinfo  {journal} {Phys. Rev. Lett.}\ }\textbf {\bibinfo
  {volume} {104}},\ \bibinfo {pages} {041301} (\bibinfo {year}
  {2010})}\BibitemShut {NoStop}%
\bibitem [{\citenamefont {Caldwell}\ \emph {et~al.}(2017)\citenamefont
  {Caldwell}, \citenamefont {Dvali}, \citenamefont {Majorovits}, \citenamefont
  {Millar}, \citenamefont {Raffelt}, \citenamefont {Redondo}, \citenamefont
  {Reimann}, \citenamefont {Simon},\ and\ \citenamefont {Steffen}}]{MADMAX}%
  \BibitemOpen
  \bibfield  {author} {\bibinfo {author} {\bibfnamefont {A.}~\bibnamefont
  {Caldwell}}, \bibinfo {author} {\bibfnamefont {G.}~\bibnamefont {Dvali}},
  \bibinfo {author} {\bibfnamefont {B.}~\bibnamefont {Majorovits}}, \bibinfo
  {author} {\bibfnamefont {A.}~\bibnamefont {Millar}}, \bibinfo {author}
  {\bibfnamefont {G.}~\bibnamefont {Raffelt}}, \bibinfo {author} {\bibfnamefont
  {J.}~\bibnamefont {Redondo}}, \bibinfo {author} {\bibfnamefont
  {O.}~\bibnamefont {Reimann}}, \bibinfo {author} {\bibfnamefont
  {F.}~\bibnamefont {Simon}}, \ and\ \bibinfo {author} {\bibfnamefont
  {F.}~\bibnamefont {Steffen}} (\bibinfo {collaboration} {MADMAX Working
  Group}),\ }\href {\doibase 10.1103/PhysRevLett.118.091801} {\bibfield
  {journal} {\bibinfo  {journal} {Phys. Rev. Lett.}\ }\textbf {\bibinfo
  {volume} {118}},\ \bibinfo {pages} {091801} (\bibinfo {year}
  {2017})}\BibitemShut {NoStop}%
\bibitem [{\citenamefont {Brubaker}\ \emph {et~al.}(2017)\citenamefont
  {Brubaker}, \citenamefont {Zhong}, \citenamefont {Gurevich}, \citenamefont
  {Cahn}, \citenamefont {Lamoreaux}, \citenamefont {Simanovskaia},
  \citenamefont {Root}, \citenamefont {Lewis}, \citenamefont {Al~Kenany},
  \citenamefont {Backes}, \citenamefont {Urdinaran}, \citenamefont {Rapidis},
  \citenamefont {Shokair}, \citenamefont {van Bibber}, \citenamefont {Palken},
  \citenamefont {Malnou}, \citenamefont {Kindel}, \citenamefont {Anil},
  \citenamefont {Lehnert},\ and\ \citenamefont {Carosi}}]{YaleAxion}%
  \BibitemOpen
  \bibfield  {author} {\bibinfo {author} {\bibfnamefont {B.~M.}\ \bibnamefont
  {Brubaker}}, \bibinfo {author} {\bibfnamefont {L.}~\bibnamefont {Zhong}},
  \bibinfo {author} {\bibfnamefont {Y.~V.}\ \bibnamefont {Gurevich}}, \bibinfo
  {author} {\bibfnamefont {S.~B.}\ \bibnamefont {Cahn}}, \bibinfo {author}
  {\bibfnamefont {S.~K.}\ \bibnamefont {Lamoreaux}}, \bibinfo {author}
  {\bibfnamefont {M.}~\bibnamefont {Simanovskaia}}, \bibinfo {author}
  {\bibfnamefont {J.~R.}\ \bibnamefont {Root}}, \bibinfo {author}
  {\bibfnamefont {S.~M.}\ \bibnamefont {Lewis}}, \bibinfo {author}
  {\bibfnamefont {S.}~\bibnamefont {Al~Kenany}}, \bibinfo {author}
  {\bibfnamefont {K.~M.}\ \bibnamefont {Backes}}, \bibinfo {author}
  {\bibfnamefont {I.}~\bibnamefont {Urdinaran}}, \bibinfo {author}
  {\bibfnamefont {N.~M.}\ \bibnamefont {Rapidis}}, \bibinfo {author}
  {\bibfnamefont {T.~M.}\ \bibnamefont {Shokair}}, \bibinfo {author}
  {\bibfnamefont {K.~A.}\ \bibnamefont {van Bibber}}, \bibinfo {author}
  {\bibfnamefont {D.~A.}\ \bibnamefont {Palken}}, \bibinfo {author}
  {\bibfnamefont {M.}~\bibnamefont {Malnou}}, \bibinfo {author} {\bibfnamefont
  {W.~F.}\ \bibnamefont {Kindel}}, \bibinfo {author} {\bibfnamefont {M.~A.}\
  \bibnamefont {Anil}}, \bibinfo {author} {\bibfnamefont {K.~W.}\ \bibnamefont
  {Lehnert}}, \ and\ \bibinfo {author} {\bibfnamefont {G.}~\bibnamefont
  {Carosi}},\ }\href {\doibase 10.1103/PhysRevLett.118.061302} {\bibfield
  {journal} {\bibinfo  {journal} {Phys. Rev. Lett.}\ }\textbf {\bibinfo
  {volume} {118}},\ \bibinfo {pages} {061302} (\bibinfo {year}
  {2017})}\BibitemShut {NoStop}%
\bibitem [{\citenamefont {Chung}(2016)}]{Cultask}%
  \BibitemOpen
  \bibfield  {author} {\bibinfo {author} {\bibfnamefont {W.}~\bibnamefont
  {Chung}},\ }\bibfield  {booktitle} {\emph {\bibinfo {booktitle}
  {{Proceedings, 15th Hellenic School and Workshops on Elementary Particle
  Physics and Gravity (CORFU2015): Corfu, Greece, September 1-25, 2015}}},\
  }\href@noop {} {\bibfield  {journal} {\bibinfo  {journal} {PoS}\ }\textbf
  {\bibinfo {volume} {CORFU2015}},\ \bibinfo {pages} {047} (\bibinfo {year}
  {2016})}\BibitemShut {NoStop}%
%%CITATION = POSCI,CORFU2015,047;%%
\bibitem [{\citenamefont {McAllister}\ \emph {et~al.}(2017)\citenamefont
  {McAllister}, \citenamefont {Flower}, \citenamefont {Ivanov}, \citenamefont
  {Goryachev}, \citenamefont {Bourhill},\ and\ \citenamefont {Tobar}}]{ORGAN}%
  \BibitemOpen
  \bibfield  {author} {\bibinfo {author} {\bibfnamefont {B.~T.}\ \bibnamefont
  {McAllister}}, \bibinfo {author} {\bibfnamefont {G.}~\bibnamefont {Flower}},
  \bibinfo {author} {\bibfnamefont {E.~N.}\ \bibnamefont {Ivanov}}, \bibinfo
  {author} {\bibfnamefont {M.}~\bibnamefont {Goryachev}}, \bibinfo {author}
  {\bibfnamefont {J.}~\bibnamefont {Bourhill}}, \ and\ \bibinfo {author}
  {\bibfnamefont {M.~E.}\ \bibnamefont {Tobar}},\ }\href {\doibase
  https://doi.org/10.1016/j.dark.2017.09.010} {\bibfield  {journal} {\bibinfo
  {journal} {Physics of the Dark Universe}\ }\textbf {\bibinfo {volume} {18}},\
  \bibinfo {pages} {67 } (\bibinfo {year} {2017})}\BibitemShut {NoStop}%
\bibitem [{\citenamefont {McAllister}\ \emph {et~al.}(2018)\citenamefont
  {McAllister}, \citenamefont {Flower}, \citenamefont {Tobar},\ and\
  \citenamefont {Tobar}}]{DielectricPaper}%
  \BibitemOpen
  \bibfield  {author} {\bibinfo {author} {\bibfnamefont {B.~T.}\ \bibnamefont
  {McAllister}}, \bibinfo {author} {\bibfnamefont {G.}~\bibnamefont {Flower}},
  \bibinfo {author} {\bibfnamefont {L.~E.}\ \bibnamefont {Tobar}}, \ and\
  \bibinfo {author} {\bibfnamefont {M.~E.}\ \bibnamefont {Tobar}},\ }\href
  {\doibase 10.1103/PhysRevApplied.9.014028} {\bibfield  {journal} {\bibinfo
  {journal} {Phys. Rev. Applied}\ }\textbf {\bibinfo {volume} {9}},\ \bibinfo
  {pages} {014028} (\bibinfo {year} {2018})}\BibitemShut {NoStop}%
\bibitem [{\citenamefont {Sikivie}\ \emph {et~al.}(2014)\citenamefont
  {Sikivie}, \citenamefont {Sullivan},\ and\ \citenamefont
  {Tanner}}]{Sikivie2014a}%
  \BibitemOpen
  \bibfield  {author} {\bibinfo {author} {\bibfnamefont {P.}~\bibnamefont
  {Sikivie}}, \bibinfo {author} {\bibfnamefont {N.}~\bibnamefont {Sullivan}}, \
  and\ \bibinfo {author} {\bibfnamefont {D.~B.}\ \bibnamefont {Tanner}},\
  }\href {\doibase 10.1103/PhysRevLett.112.131301} {\bibfield  {journal}
  {\bibinfo  {journal} {Phys. Rev. Lett.}\ }\textbf {\bibinfo {volume} {112}},\
  \bibinfo {pages} {131301} (\bibinfo {year} {2014})}\BibitemShut {NoStop}%
\bibitem [{\citenamefont {Hoskins}\ \emph {et~al.}(2011)\citenamefont
  {Hoskins}, \citenamefont {Hwang}, \citenamefont {Martin}, \citenamefont
  {Sikivie}, \citenamefont {Sullivan}, \citenamefont {Tanner}, \citenamefont
  {Hotz}, \citenamefont {Rosenberg}, \citenamefont {Rybka}, \citenamefont
  {Wagner}, \citenamefont {Asztalos}, \citenamefont {Carosi}, \citenamefont
  {Hagmann}, \citenamefont {Kinion}, \citenamefont {van Bibber}, \citenamefont
  {Bradley},\ and\ \citenamefont {Clarke}}]{ADMX2011}%
  \BibitemOpen
  \bibfield  {author} {\bibinfo {author} {\bibfnamefont {J.}~\bibnamefont
  {Hoskins}}, \bibinfo {author} {\bibfnamefont {J.}~\bibnamefont {Hwang}},
  \bibinfo {author} {\bibfnamefont {C.}~\bibnamefont {Martin}}, \bibinfo
  {author} {\bibfnamefont {P.}~\bibnamefont {Sikivie}}, \bibinfo {author}
  {\bibfnamefont {N.~S.}\ \bibnamefont {Sullivan}}, \bibinfo {author}
  {\bibfnamefont {D.~B.}\ \bibnamefont {Tanner}}, \bibinfo {author}
  {\bibfnamefont {M.}~\bibnamefont {Hotz}}, \bibinfo {author} {\bibfnamefont
  {L.~J.}\ \bibnamefont {Rosenberg}}, \bibinfo {author} {\bibfnamefont
  {G.}~\bibnamefont {Rybka}}, \bibinfo {author} {\bibfnamefont
  {A.}~\bibnamefont {Wagner}}, \bibinfo {author} {\bibfnamefont {S.~J.}\
  \bibnamefont {Asztalos}}, \bibinfo {author} {\bibfnamefont {G.}~\bibnamefont
  {Carosi}}, \bibinfo {author} {\bibfnamefont {C.}~\bibnamefont {Hagmann}},
  \bibinfo {author} {\bibfnamefont {D.}~\bibnamefont {Kinion}}, \bibinfo
  {author} {\bibfnamefont {K.}~\bibnamefont {van Bibber}}, \bibinfo {author}
  {\bibfnamefont {R.}~\bibnamefont {Bradley}}, \ and\ \bibinfo {author}
  {\bibfnamefont {J.}~\bibnamefont {Clarke}},\ }\href {\doibase
  10.1103/PhysRevD.84.121302} {\bibfield  {journal} {\bibinfo  {journal} {Phys.
  Rev. D}\ }\textbf {\bibinfo {volume} {84}},\ \bibinfo {pages} {121302}
  (\bibinfo {year} {2011})}\BibitemShut {NoStop}%
\bibitem [{\citenamefont {McAllister}\ \emph
  {et~al.}(2016{\natexlab{a}})\citenamefont {McAllister}, \citenamefont
  {Parker},\ and\ \citenamefont {Tobar}}]{McAllisterFormFactor}%
  \BibitemOpen
  \bibfield  {author} {\bibinfo {author} {\bibfnamefont {B.~T.}\ \bibnamefont
  {McAllister}}, \bibinfo {author} {\bibfnamefont {S.~R.}\ \bibnamefont
  {Parker}}, \ and\ \bibinfo {author} {\bibfnamefont {M.~E.}\ \bibnamefont
  {Tobar}},\ }\href {\doibase 10.1103/PhysRevLett.117.159901,
  10.1103/PhysRevLett.116.161804} {\bibfield  {journal} {\bibinfo  {journal}
  {Phys. Rev. Lett.}\ }\textbf {\bibinfo {volume} {116}},\ \bibinfo {pages}
  {161804} (\bibinfo {year} {2016}{\natexlab{a}})},\ \bibinfo {note} {[Erratum:
  Phys. Rev. Lett.117,no.15,159901(2016)]},\ \Eprint
  {http://arxiv.org/abs/1607.01928} {arXiv:1607.01928 [hep-ph]} \BibitemShut
  {NoStop}%
%%CITATION = ARXIV:1607.01928;%%
\bibitem [{\citenamefont {Garcon}\ \emph {et~al.}(2018)\citenamefont {Garcon},
  \citenamefont {Aybas}, \citenamefont {Blanchard}, \citenamefont {Centers},
  \citenamefont {Figueroa}, \citenamefont {Graham}, \citenamefont {Kimball},
  \citenamefont {Rajendran}, \citenamefont {Sendra}, \citenamefont {Sushkov},
  \citenamefont {Trahms}, \citenamefont {Wang}, \citenamefont {Wickenbrock},
  \citenamefont {Wu},\ and\ \citenamefont {Budker}}]{Garcon}%
  \BibitemOpen
  \bibfield  {author} {\bibinfo {author} {\bibfnamefont {A.}~\bibnamefont
  {Garcon}}, \bibinfo {author} {\bibfnamefont {D.}~\bibnamefont {Aybas}},
  \bibinfo {author} {\bibfnamefont {J.~W.}\ \bibnamefont {Blanchard}}, \bibinfo
  {author} {\bibfnamefont {G.}~\bibnamefont {Centers}}, \bibinfo {author}
  {\bibfnamefont {N.~L.}\ \bibnamefont {Figueroa}}, \bibinfo {author}
  {\bibfnamefont {P.~W.}\ \bibnamefont {Graham}}, \bibinfo {author}
  {\bibfnamefont {D.~F.~J.}\ \bibnamefont {Kimball}}, \bibinfo {author}
  {\bibfnamefont {S.}~\bibnamefont {Rajendran}}, \bibinfo {author}
  {\bibfnamefont {M.~G.}\ \bibnamefont {Sendra}}, \bibinfo {author}
  {\bibfnamefont {A.~O.}\ \bibnamefont {Sushkov}}, \bibinfo {author}
  {\bibfnamefont {L.}~\bibnamefont {Trahms}}, \bibinfo {author} {\bibfnamefont
  {T.}~\bibnamefont {Wang}}, \bibinfo {author} {\bibfnamefont {A.}~\bibnamefont
  {Wickenbrock}}, \bibinfo {author} {\bibfnamefont {T.}~\bibnamefont {Wu}}, \
  and\ \bibinfo {author} {\bibfnamefont {D.}~\bibnamefont {Budker}},\ }\href
  {http://stacks.iop.org/2058-9565/3/i=1/a=014008} {\bibfield  {journal}
  {\bibinfo  {journal} {Quantum Science and Technology}\ }\textbf {\bibinfo
  {volume} {3}},\ \bibinfo {pages} {014008} (\bibinfo {year}
  {2018})}\BibitemShut {NoStop}%
\bibitem [{\citenamefont {Budker}\ \emph {et~al.}(2014)\citenamefont {Budker},
  \citenamefont {Graham}, \citenamefont {Ledbetter}, \citenamefont
  {Rajendran},\ and\ \citenamefont {Sushkov}}]{BudkerPRX}%
  \BibitemOpen
  \bibfield  {author} {\bibinfo {author} {\bibfnamefont {D.}~\bibnamefont
  {Budker}}, \bibinfo {author} {\bibfnamefont {P.~W.}\ \bibnamefont {Graham}},
  \bibinfo {author} {\bibfnamefont {M.}~\bibnamefont {Ledbetter}}, \bibinfo
  {author} {\bibfnamefont {S.}~\bibnamefont {Rajendran}}, \ and\ \bibinfo
  {author} {\bibfnamefont {A.~O.}\ \bibnamefont {Sushkov}},\ }\href {\doibase
  10.1103/PhysRevX.4.021030} {\bibfield  {journal} {\bibinfo  {journal} {Phys.
  Rev. X}\ }\textbf {\bibinfo {volume} {4}},\ \bibinfo {pages} {021030}
  (\bibinfo {year} {2014})}\BibitemShut {NoStop}%
\bibitem [{\citenamefont {Kim}\ \emph {et~al.}(2014)\citenamefont {Kim},
  \citenamefont {Semertzidis},\ and\ \citenamefont {Tsujikawa}}]{Kim2014}%
  \BibitemOpen
  \bibfield  {author} {\bibinfo {author} {\bibfnamefont {J.~E.}\ \bibnamefont
  {Kim}}, \bibinfo {author} {\bibfnamefont {Y.~K.}\ \bibnamefont
  {Semertzidis}}, \ and\ \bibinfo {author} {\bibfnamefont {S.}~\bibnamefont
  {Tsujikawa}},\ }\href {\doibase 10.3389/fphy.2014.00060} {\bibfield
  {journal} {\bibinfo  {journal} {Frontiers in Physics}\ }\textbf {\bibinfo
  {volume} {2}},\ \bibinfo {pages} {60} (\bibinfo {year} {2014})}\BibitemShut
  {NoStop}%
\bibitem [{\citenamefont {Ko}\ \emph {et~al.}(2016)\citenamefont {Ko},
  \citenamefont {Themann}, \citenamefont {Jang}, \citenamefont {Choi},
  \citenamefont {Kim}, \citenamefont {Lee}, \citenamefont {Lee}, \citenamefont
  {Won},\ and\ \citenamefont {Semertzidis}}]{PhysRevD.94.111702}%
  \BibitemOpen
  \bibfield  {author} {\bibinfo {author} {\bibfnamefont {B.~R.}\ \bibnamefont
  {Ko}}, \bibinfo {author} {\bibfnamefont {H.}~\bibnamefont {Themann}},
  \bibinfo {author} {\bibfnamefont {W.}~\bibnamefont {Jang}}, \bibinfo {author}
  {\bibfnamefont {J.}~\bibnamefont {Choi}}, \bibinfo {author} {\bibfnamefont
  {D.}~\bibnamefont {Kim}}, \bibinfo {author} {\bibfnamefont {M.~J.}\
  \bibnamefont {Lee}}, \bibinfo {author} {\bibfnamefont {J.}~\bibnamefont
  {Lee}}, \bibinfo {author} {\bibfnamefont {E.}~\bibnamefont {Won}}, \ and\
  \bibinfo {author} {\bibfnamefont {Y.~K.}\ \bibnamefont {Semertzidis}},\
  }\href {\doibase 10.1103/PhysRevD.94.111702} {\bibfield  {journal} {\bibinfo
  {journal} {Phys. Rev. D}\ }\textbf {\bibinfo {volume} {94}},\ \bibinfo
  {pages} {111702} (\bibinfo {year} {2016})}\BibitemShut {NoStop}%
\bibitem [{\citenamefont {Hoang}\ \emph {et~al.}(2017)\citenamefont {Hoang},
  \citenamefont {Jeong}, \citenamefont {Cao}, \citenamefont {Shin},
  \citenamefont {Ko},\ and\ \citenamefont {Semertzidis}}]{HOANG2017}%
  \BibitemOpen
  \bibfield  {author} {\bibinfo {author} {\bibfnamefont {P.~L.}\ \bibnamefont
  {Hoang}}, \bibinfo {author} {\bibfnamefont {J.}~\bibnamefont {Jeong}},
  \bibinfo {author} {\bibfnamefont {B.~X.}\ \bibnamefont {Cao}}, \bibinfo
  {author} {\bibfnamefont {Y.}~\bibnamefont {Shin}}, \bibinfo {author}
  {\bibfnamefont {B.}~\bibnamefont {Ko}}, \ and\ \bibinfo {author}
  {\bibfnamefont {Y.}~\bibnamefont {Semertzidis}},\ }\href {\doibase
  https://doi.org/10.1016/j.dark.2017.04.004} {\bibfield  {journal} {\bibinfo
  {journal} {Physics of the Dark Universe}\ } (\bibinfo {year} {2017}),\
  https://doi.org/10.1016/j.dark.2017.04.004}\BibitemShut {NoStop}%
\bibitem [{\citenamefont {McAllister}\ \emph
  {et~al.}(2016{\natexlab{b}})\citenamefont {McAllister}, \citenamefont
  {Parker},\ and\ \citenamefont {Tobar}}]{McAllister2016fux}%
  \BibitemOpen
  \bibfield  {author} {\bibinfo {author} {\bibfnamefont {B.~T.}\ \bibnamefont
  {McAllister}}, \bibinfo {author} {\bibfnamefont {S.~R.}\ \bibnamefont
  {Parker}}, \ and\ \bibinfo {author} {\bibfnamefont {M.~E.}\ \bibnamefont
  {Tobar}},\ }\href {\doibase 10.1103/PhysRevD.94.042001} {\bibfield  {journal}
  {\bibinfo  {journal} {Phys. Rev.}\ }\textbf {\bibinfo {volume} {D94}},\
  \bibinfo {pages} {042001} (\bibinfo {year} {2016}{\natexlab{b}})},\ \Eprint
  {http://arxiv.org/abs/1605.05427} {arXiv:1605.05427 [physics.ins-det]}
  \BibitemShut {NoStop}%
%%CITATION = ARXIV:1605.05427;%%
\bibitem [{\citenamefont {Hong}\ \emph {et~al.}(2014)\citenamefont {Hong},
  \citenamefont {Kim}, \citenamefont {Nam},\ and\ \citenamefont
  {Semertzidis}}]{Jooyoo2014}%
  \BibitemOpen
  \bibfield  {author} {\bibinfo {author} {\bibfnamefont {J.}~\bibnamefont
  {Hong}}, \bibinfo {author} {\bibfnamefont {J.~E.}\ \bibnamefont {Kim}},
  \bibinfo {author} {\bibfnamefont {S.}~\bibnamefont {Nam}}, \ and\ \bibinfo
  {author} {\bibfnamefont {Y.}~\bibnamefont {Semertzidis}},\ }\href@noop {}
  {\bibfield  {journal} {\bibinfo  {journal} {arXiv:1403.1576 [hep-ph]}\ }
  (\bibinfo {year} {2014})}\BibitemShut {NoStop}%
\bibitem [{\citenamefont {Kahn}\ \emph {et~al.}(2016)\citenamefont {Kahn},
  \citenamefont {Safdi},\ and\ \citenamefont {Thaler}}]{ABRACADABRA}%
  \BibitemOpen
  \bibfield  {author} {\bibinfo {author} {\bibfnamefont {Y.}~\bibnamefont
  {Kahn}}, \bibinfo {author} {\bibfnamefont {B.~R.}\ \bibnamefont {Safdi}}, \
  and\ \bibinfo {author} {\bibfnamefont {J.}~\bibnamefont {Thaler}},\ }\href
  {\doibase 10.1103/PhysRevLett.117.141801} {\bibfield  {journal} {\bibinfo
  {journal} {Phys. Rev. Lett.}\ }\textbf {\bibinfo {volume} {117}},\ \bibinfo
  {pages} {141801} (\bibinfo {year} {2016})},\ \Eprint
  {http://arxiv.org/abs/1602.01086} {arXiv:1602.01086 [hep-ph]} \BibitemShut
  {NoStop}%
%%CITATION = ARXIV:1602.01086;%%
\bibitem [{\citenamefont {Hong}\ \emph {et~al.}(1990)\citenamefont {Hong},
  \citenamefont {Kim},\ and\ \citenamefont {Sikivie}}]{Jooyoo1990}%
  \BibitemOpen
  \bibfield  {author} {\bibinfo {author} {\bibfnamefont {J.}~\bibnamefont
  {Hong}}, \bibinfo {author} {\bibfnamefont {J.~E.}\ \bibnamefont {Kim}}, \
  and\ \bibinfo {author} {\bibfnamefont {P.}~\bibnamefont {Sikivie}},\ }\href
  {\doibase 10.1103/PhysRevD.42.1847} {\bibfield  {journal} {\bibinfo
  {journal} {Phys. Rev. D}\ }\textbf {\bibinfo {volume} {42}},\ \bibinfo
  {pages} {1847} (\bibinfo {year} {1990})}\BibitemShut {NoStop}%
\bibitem [{\citenamefont {Hong}\ and\ \citenamefont {Kim}(1991)}]{HONG1991197}%
  \BibitemOpen
  \bibfield  {author} {\bibinfo {author} {\bibfnamefont {J.}~\bibnamefont
  {Hong}}\ and\ \bibinfo {author} {\bibfnamefont {J.~E.}\ \bibnamefont {Kim}},\
  }\href {\doibase https://doi.org/10.1016/0370-2693(91)90040-W} {\bibfield
  {journal} {\bibinfo  {journal} {Physics Letters B}\ }\textbf {\bibinfo
  {volume} {265}},\ \bibinfo {pages} {197 } (\bibinfo {year}
  {1991})}\BibitemShut {NoStop}%
\bibitem [{\citenamefont {Tobar}\ \emph {et~al.}(2018)\citenamefont {Tobar},
  \citenamefont {McAllister},\ and\ \citenamefont {Goryachev}}]{MTAxEd}%
  \BibitemOpen
  \bibfield  {author} {\bibinfo {author} {\bibfnamefont {M.~E.}\ \bibnamefont
  {Tobar}}, \bibinfo {author} {\bibfnamefont {B.~T.}\ \bibnamefont
  {McAllister}}, \ and\ \bibinfo {author} {\bibfnamefont {M.}~\bibnamefont
  {Goryachev}},\ }\href@noop {} {\bibfield  {journal} {\bibinfo  {journal}
  {arXiv:1809.01654 [hep-ph]}\ } (\bibinfo {year} {2018})}\BibitemShut {NoStop}%
\bibitem [{\citenamefont {Wilczek}(1987)}]{Wilczek:1987aa}%
  \BibitemOpen
  \bibfield  {author} {\bibinfo {author} {\bibfnamefont {F.}~\bibnamefont
  {Wilczek}},\ }\href {http://link.aps.org/doi/10.1103/PhysRevLett.58.1799}
  {\bibfield  {journal} {\bibinfo  {journal} {Physical Review Letters}\
  }\textbf {\bibinfo {volume} {58}},\ \bibinfo {pages} {1799} (\bibinfo {year}
  {1987})}\BibitemShut {NoStop}%
\bibitem [{\citenamefont {Daw}(1998)}]{Daw:1998jm}%
  \BibitemOpen
  \bibfield  {author} {\bibinfo {author} {\bibfnamefont {E.~J.}\ \bibnamefont
  {Daw}},\ }\emph {\bibinfo {title} {{A search for halo axions}}},\ \href
  {http://wwwlib.umi.com/dissertations/fullcit?334417} {Ph.D. thesis},\
  \bibinfo  {school} {MIT} (\bibinfo {year} {1998})\BibitemShut {NoStop}%
%%CITATION = UMI-33-4417;%%
\bibitem [{\citenamefont {Goryachev}\ \emph {et~al.}(2014)\citenamefont
  {Goryachev}, \citenamefont {Ivanov}, \citenamefont {van Kann}, \citenamefont
  {Galliou},\ and\ \citenamefont {Tobar}}]{SQUIDQuartz}%
  \BibitemOpen
  \bibfield  {author} {\bibinfo {author} {\bibfnamefont {M.}~\bibnamefont
  {Goryachev}}, \bibinfo {author} {\bibfnamefont {E.~N.}\ \bibnamefont
  {Ivanov}}, \bibinfo {author} {\bibfnamefont {F.}~\bibnamefont {van Kann}},
  \bibinfo {author} {\bibfnamefont {S.}~\bibnamefont {Galliou}}, \ and\
  \bibinfo {author} {\bibfnamefont {M.~E.}\ \bibnamefont {Tobar}},\ }\href
  {\doibase 10.1063/1.4898813} {\bibfield  {journal} {\bibinfo  {journal}
  {Applied Physics Letters}\ }\textbf {\bibinfo {volume} {105}},\ \bibinfo
  {pages} {153505} (\bibinfo {year} {2014})},\ \Eprint
  {http://arxiv.org/abs/https://doi.org/10.1063/1.4898813}
  {https://doi.org/10.1063/1.4898813} \BibitemShut {NoStop}%
\bibitem [{\citenamefont {Fairbairn}\ \emph {et~al.}(2018)\citenamefont
  {Fairbairn}, \citenamefont {Marsh}, \citenamefont {Quevillon},\ and\
  \citenamefont {Rozier}}]{AxionMC}%
  \BibitemOpen
  \bibfield  {author} {\bibinfo {author} {\bibfnamefont {M.}~\bibnamefont
  {Fairbairn}}, \bibinfo {author} {\bibfnamefont {D.~J.~E.}\ \bibnamefont
  {Marsh}}, \bibinfo {author} {\bibfnamefont {J.}~\bibnamefont {Quevillon}}, \
  and\ \bibinfo {author} {\bibfnamefont {S.}~\bibnamefont {Rozier}},\ }\href
  {\doibase 10.1103/PhysRevD.97.083502} {\bibfield  {journal} {\bibinfo
  {journal} {Phys. Rev. D}\ }\textbf {\bibinfo {volume} {97}},\ \bibinfo
  {pages} {083502} (\bibinfo {year} {2018})}\BibitemShut {NoStop}%
\bibitem [{\citenamefont {Zioutas}\ \emph {et~al.}(2017)\citenamefont {Zioutas}
  \emph {et~al.}}]{StreamingDM}%
  \BibitemOpen
  \bibfield  {author} {\bibinfo {author} {\bibfnamefont {K.}~\bibnamefont
  {Zioutas}} \emph {et~al.},\ }\href@noop {} {\  (\bibinfo {year} {2017})},\
  \Eprint {http://arxiv.org/abs/1703.01436} {arXiv:1703.01436
  [physics.ins-det]} \BibitemShut {NoStop}%
%%CITATION = ARXIV:1703.01436;%%
\bibitem [{\citenamefont {Kosteleck{\'y}}\ and\ \citenamefont
  {Mewes}(2002)}]{KM}%
  \BibitemOpen
  \bibfield  {author} {\bibinfo {author} {\bibfnamefont {V.~A.}\ \bibnamefont
  {Kosteleck{\'y}}}\ and\ \bibinfo {author} {\bibfnamefont {M.}~\bibnamefont
  {Mewes}},\ }\href {http://link.aps.org/doi/10.1103/PhysRevD.66.056005}
  {\bibfield  {journal} {\bibinfo  {journal} {Phys. Rev. D}\ }\textbf {\bibinfo
  {volume} {66}},\ \bibinfo {pages} {056005} (\bibinfo {year}
  {2002})}\BibitemShut {NoStop}%
\bibitem [{HIA(2016)}]{HIA}%
  \BibitemOpen
  \href@noop {} {\emph {\bibinfo {title} {HFC 50 D / E Dual Cryogenic Ultra Low
  Noise RF-Amplifier}}},\ \bibinfo {organization} {Stahl Electronics} (\bibinfo
  {year} {2016}),\ \bibinfo {note} {version 2.38}\BibitemShut {NoStop}%
\end{thebibliography}
\end{document}